\global\def\draftcontrol{0}

   \def\versionno{Lectures on Lambda Positive Universe -- draft   }

\catcode`\@=11

\expandafter\ifx\csname draftcontrol\endcsname\relax\global\def\draftcontrol{0}
\fi

{\count255=\time\divide\count255 by 60
\xdef\hourmin{\number\count255}
\multiply\count255 by-60\advance\count255 by\time
\xdef\hourmin{\hourmin:\ifnum\count255<10 0\fi\the\count255}}
\def\draftdate{\number\month/\number\day/\number\year\ \ \ \hourmin }

\newcommand\makepapertitle{\par
  \begingroup
    \renewcommand\thefootnote{\@fnsymbol\c@footnote}%
    \def\@makefnmark{\rlap{\@textsuperscript{\normalfont\@thefnmark}}}%
    \long\def\@makefntext##1{\parindent 1em\noindent
            \hb@xt@1.8em{%
                \hss\@textsuperscript{\normalfont\@thefnmark}}##1}%
     \newpage
     \global\@topnum\z@   
     \@makepapertitle
     \thispagestyle{empty}\@thanks
  \endgroup
  \setcounter{footnote}{0}%
  \global\let\thanks\relax
  \global\let\makepapertitle\relax
  \global\let\@makepapertitle\relax
  \global\let\@thanks\@empty
  \global\let\@author\@empty
  \global\let\@date\@empty
  \global\let\@title\@empty
  \global\let\title\relax
  \global\let\author\relax
  \global\let\date\relax
  \global\let\and\relax
  \def\version{\let\version\@version\@gobble}
}
\def\@makepapertitle{%
  \newpage
   \ifnum\draftcontrol=1 {}
   \version\versionno
   \vskip 3em%
   \else
   \hfill\hbox to 3cm {\parbox{4cm}{\@pubnum}\hss}%
   \vskip 3em%
   \fi
   \begin{center}%
   \let \footnote \thanks
     {\LARGE {\@title}}%
     \vskip 1.5em%
     {\normalsize
       \lineskip .5em%
       \begin{tabular}[t]{l}%
         \@author
       \end{tabular}\par}%
     \vskip 1.5em%
     {\@bstract}%
     \end{center}%
     \vskip 1.5em
     \@date%
   \par
}

\gdef\@pubnum{}
\def\pubnum#1{%
  \gdef\@pubnum{#1}}

\gdef\@bstract{}
\def\Abstract#1{%
  \gdef\@bstract{%
   \parbox{\textwidth-0pc}{%
   \centerline{\bf Abstract}\penalty1000%
\kern.2cm%
\noindent
\renewcommand\baselinestretch{1.0}%
{#1}}}
}

\def\ps@paper{\let\@mkboth\@gobbletwo%
     \ifnum\draftcontrol=1
    \def\@oddfoot{\hbox to \textwidth{\tiny \versionno \hfil\tiny\draftdate}%
    \hskip -\textwidth \hbox to \textwidth{\hfil\rm\thepage\hfil}}%
     \else\def\@oddfoot{\hbox to \textwidth{\hfil\rm\thepage\hfil}}
     \fi
     \let\@evenfoot\@oddfoot
}

\def\body{\clearpage
          \pagestyle{paper}
    }
\newenvironment{acknowledgements}{%
\vskip 3.25ex
\noindent {\bf Acknowledgements}
}

\def\@version#1{\ifnum\draftcontrol=1
\typeout{}\typeout{#1}\typeout{}
\vskip3mm\centerline{\hbox{\fbox{\normalsize{\tt DRAFT -- #1 -- }
                   {\draftdate}}}}\vskip3mm
\fi}
\let\version\@version
\long\def\eqlabel#1{\ifnum\draftcontrol=1
                    \tag@false  
                    \tag*{(\theequation) \hbox to -0.2cm{\hspace{0cm}\small{#1}\hss}}
                    \refstepcounter{equation}
                    \edef\@currentlabel{\theequation}
                    \ltx@label{#1}          
                    \else
                    \label{#1}
                    \fi
                    }
\let\st@bibitem\@bibitem
\let\st@lbibitem\@lbibitem
\ifnum\draftcontrol=1
  \def\@bibitem#1{%
    \st@bibitem{#1}\a@@label{#1}\ignorespaces}
  \def\@lbibitem[#1]#2{%
    \st@lbibitem[#1]{#2}\a@@label{#2}\ignorespaces}
  \def\a@@label#1{%
    \gdef\a@lab{\smash{\normalfont\small#1}}
    \ifvmode
      \if@inlabel
        \global\setbox\@labels\hbox{%
          \llap{\a@lab\let\a@lab\relax
                \kern\@totalleftmargin\kern\marginparsep}%
          \box\@labels}%
      \fi
    \fi}
\fi

\documentclass[10pt]{article}

\usepackage{amsmath,amssymb,array,calc,epsfig,bm}
\usepackage{graphicx}        
\usepackage{multicol}        

\usepackage{amsmath,amssymb,bm,epsfig}

\usepackage{cite}
\usepackage[
      colorlinks=true,
      linkcolor=red,  
      urlcolor=blue,    
      filecolor=red,     
      linkcolor=blue,
      citecolor = red,
      pdfstartview=FitV,
      pdftitle={},
        pdfauthor={Paolo Benincasa},
        pdfsubject={},
        pdfkeywords={},
        pdfpagemode=None,
        bookmarksopen=true    
      ]{hyperref}

\usepackage{centernot}      
\usepackage{mathpazo}
\usepackage{stmaryrd}
\usepackage{pifont}
\usepackage{colortbl}
\usepackage{multirow,bigdelim}
\usepackage{arydshln}
\usepackage{tikz, pict2e}
\usepackage{geometry}
\usepackage{lmodern}
\usepackage{wrapfig}
\usepackage{float}
\usepackage{enumitem}
\usepackage[framemethod=default]{mdframed}
\usepackage{array}
\usepackage{mathrsfs}

\usepackage{placeins}
\usepackage{circuitikz}

\usepackage[T1]{fontenc}
\usepackage[utf8]{inputenc}
\usepackage[english]{babel}

\PassOptionsToPackage{table}{xcolor}

\usepackage{xspace}
\usepackage{ifthen}
\usepackage{epstopdf}
\usepackage{amsthm}

\usepackage{relsize}
\usepackage{tabularx}
\usepackage{booktabs}
\usepackage{soul}
\usepackage{pdfpages}
\usepackage[most]{tcolorbox}
\usepackage{capt-of}
\usepackage{xfrac}
\usepackage{empheq}
\usepackage{slashed}
\usepackage{cancel}
\usepackage{upgreek}
\usepackage[skip=1em]{parskip}

\usepackage{latexsym}
\usepackage{ifoddpage}

\usepackage{subcaption}

\usepackage{cancel}
\usepackage{comment}
\usepackage{shuffle}

\usepackage{savesym}
\savesymbol{Ref}
\savesymbol{erf}
\usepackage{physics}
\restoresymbol{phys}{Ref}
\restoresymbol{phys}{erf}

\usepackage[compat=1.1.0]{tikz-feynman}
\tikzfeynmanset{compat=1.1.0}

\definecolor{myblue}{RGB}{166, 193, 222}
\definecolor{myblue2}{RGB}{68, 114, 196}
\definecolor{mygreen}{RGB}{175, 218, 162}
\definecolor{mygreen2}{RGB}{133, 210, 110}
\definecolor{myorange2}{RGB}{255, 192, 0}

\newtcolorbox{shade}[1][]{
  top=-7pt,bottom=6pt,
  sharp corners=all,
  colback=white!95!black,
  colframe=white,
}

\newcommand{\Li}{\mathrm{Li}}

\def\Disc{\mathrm{Disc}}
\def\id{\mathbb{I}} 
\newcommand{\sgn}[1]{\mathrm{sgn}( #1)}
\def\res{\mathrm{Res}}
\def\bubble{\begin{tikzpicture}[scale=0.1]
    \draw (-2, 0) -- (-1, 0);
    \draw (1, 0) -- (2, 0);
    \draw (0,0) circle (1);
\end{tikzpicture}}
\DeclareRobustCommand{\bubbletry}{%
  \begin{tikzpicture}[scale=0.1,baseline=-0.6ex]
    \draw (-2, 0) -- (-1, 0);
    \draw (1, 0) -- (2, 0);
    \draw (0,0) circle (1);
  \end{tikzpicture}%
}

\newcolumntype{P}[1]{>{\centering\arraybackslash}p{#1}}
\newcolumntype{L}{>{$}l<{$}}

\usetikzlibrary{calc,backgrounds}
\usetikzlibrary{arrows,shapes,positioning,shadows,trees}
\usetikzlibrary{arrows.meta, shapes.geometric}
\usetikzlibrary{mindmap}
\usetikzlibrary{decorations.pathreplacing}
\usetikzlibrary{decorations.pathmorphing}
\usetikzlibrary{decorations.markings}
\usetikzlibrary{matrix}
\usetikzlibrary{patterns.meta}

\tikzstyle arrowstyle=[scale=1]
\tikzstyle directed=[postaction={decorate,decoration={markings,
    mark=at position .5 with {\arrow[arrowstyle]{stealth}}}}]
\tikzstyle reverse directed=[postaction={decorate,decoration={markings,
    mark=at position .5 with {\arrowreversed[arrowstyle]{stealth};}}}]

\ifnum\draftcontrol=1
\fi

\renewcommand\baselinestretch{1.25}
\renewcommand\section{\@startsection {section}{1}{\z@}%
                                   {-3.5ex \@plus -1ex \@minus -.2ex}%
                                   {2.3ex \@plus.2ex}%
                                   {\normalfont\large\bfseries}}
\renewcommand\subsection{\@startsection{subsection}{2}{\z@}%
                                   {-3.25ex\@plus -1ex \@minus -.2ex}%
                                   {1.5ex \@plus .2ex}%
                                   {\normalfont\normalsize\bfseries}}
\renewcommand\subsubsection{\@startsection{subsubsection}{3}{\z@}%
                                   {-3.25ex\@plus -1ex \@minus -.2ex}%
                                   {1.5ex \@plus .2ex}%
                                   {\normalfont\normalsize\it}}
\renewcommand\paragraph{\@startsection{paragraph}{4}{\z@}%
                                   {-3.25ex\@plus -1ex \@minus -.2ex}%
                                   {1.5ex \@plus .2ex}%
                                   {\normalfont\normalsize\bf}}

\numberwithin{equation}{section}

\def\revise#1       {\raisebox{-0em}{\rule{3pt}{1em}}%
                     \marginpar{\raisebox{.5em}{\vrule width3pt\
                     \vrule width0pt height 0pt depth0.5em
                     \hbox to 0cm{\hspace{0cm}{%
                     \parbox[t]{4em}{\raggedright\footnotesize{#1}}}\hss}}}}

\def\sqr#1#2{{\vcenter{\vbox{\hrule height.#2pt
 \hbox{\vrule width.#2pt height#1pt \kern#1pt
 \vrule width.#2pt}\hrule height.#2pt}}}}

\catcode`\@=12

\newcommand{\ee}{e}

\newcommand{\beq}{\begin{equation}}
\newcommand{\eeq}{\end{equation}}
\newcommand{\bea}{\begin{equation}\begin{aligned}}
\newcommand{\eea}{\end{aligned}\end{equation}}

\newlength{\wsingfig}
\newlength{\wdblefig}
\newlength{\wquadfig}
\newlength{\wtriplefig}
\newcommand   \lr  [3] {\left #1 #2 \right #3}

\newcommand   \ptl {\partial}

\newcommand   \lptl{\raise .8ex\hbox{$^\leftarrow$} \hspace{-9pt} \partial}
\newcommand   \lrptl{\raise .8ex\hbox{$^\leftrightarrow$} \hspace{-9pt} \partial}

\renewcommand \a  {\alpha}

\newcommand   \g  {\gamma}

\newcommand   \de {\delta}

\newcommand   \s  {\sigma}

\renewcommand \Re    {\mathop{\mathrm{Re}}}
\renewcommand \Im    {\mathop{\mathrm{Im}}}

\newcommand   \ci    {\mathrm{i}}   

\newcommand   \be {\mathbf{e}}

\def\be#1\ee{\begin{align}#1\end{align}}

\newcounter{ssooeexercise}[section]
\renewcommand{\thessooeexercise}{\thesection.\arabic{ssooeexercise}}

\newtcolorbox{exbox}{
  colback=gray!10,
  colframe=gray!20,
  boxrule=1pt,
  arc=5pt,
  breakable,
  left=1em,
  right=1em,
  top=0.8em,
  bottom=0.8em
}

\newcommand{\newexercise}[3][]{%
  \refstepcounter{ssooeexercise}%
  \expandafter\gdef\csname ex@content@#2\endcsname{#3}%
  \expandafter\xdef\csname ex@number@#2\endcsname{\thessooeexercise}%
  \begin{exbox}%
    \textbf{Exercise~\thessooeexercise:} #3
    \ifx&#1&\else\label{#1}\fi
  \end{exbox}%
}

\newcommand{\printexercise}[1]{%
  \begin{exbox}%
    \textbf{Exercise~\csname ex@number@#1\endcsname:}
    \csname ex@content@#1\endcsname
  \end{exbox}%
}

\begin{document}


\title{\underline{\Large\bf The Disordered Universe 2025:}\\%
       \Large\bf Analytic Properties of Amplitudes and Correlators}

\pubnum{%
}

\author{%
	{\scshape Lecturer:} {\bf Andrew J. McLeod}${}^{\circledast}$, \\
    {\scshape Notes written by:} {\bf Mattia Capuano}${}^{\star}$, {\bf Chandramouli Chowdhury}${}^{\ast\,\circledcirc}$, {\bf Chantelle Esper}${}^{\oplus}$, \\
      \hspace{3.375cm} {\bf Alessandro Palazio}${}^{\star}$, {\bf Jonah Stalknecht}${}^{\ominus}$, {\bf Yong Sheng Yap}${}^{\odot,\otimes}$
    \\
    {\scshape Edited by:} {\bf Paolo Benincasa}${}^{\dagger\,\ddagger}$, {\bf Guilherme Leite Pimentel}${}^{\diamond}$
    \\ [0.4cm]
    \ttfamily ${}^{\circledast}$ Higgs Centre for Theoretical Physics, School of Physics and Astronomy, \\
    \ttfamily $\phantom{{}^{\dagger}}$ The University of Edinburgh Edinburgh EH9 3FD, Scotland, UK \\
    \ttfamily ${}^{\star}$ Department of Physics, Astronomy and Mathematics, \\
    \ttfamily $\phantom{{}^{\dagger}}$ University of Hertfordshire, Hatfield, Hertfordshire, AL10 9AB, UK \\
    \ttfamily ${}^{\ast}$ Mathematical Sciences and STAG Research Centre,  \\ 
    \ttfamily$\phantom{{}^{\dagger}}$ University of Southampton, Highfield, Southampton SO17 1BJ, UK \\
    \ttfamily ${}^{\circledcirc}$ Institute of Theoretical Physics,  \\ 
    \ttfamily$\phantom{{}^{\dagger}}$ Chinese Academy of Sciences, Beijing 100190, China \\
    \ttfamily ${}^{\oplus}$ School of Mathematics and Hamilton Mathematics Institute, \\
    \ttfamily $\phantom{{}^{\dagger}}$ Trinity College, Dublin 2, Ireland  \\
    \ttfamily ${}^{\ominus}$ Institute of Particle and Nuclear Physics, Faculty of Mathematics and Physics,\\
    \ttfamily $\phantom{{}^{\dagger}}$ Charles University, Prague 180 00, Czech Republic\\
    \ttfamily ${}^{\odot}$ Department of Applied Mathematics and Theoretical Physics, \\
    \ttfamily $\phantom{{}^{\dagger}}$ University of Cambridge,  Wilberforce Road, Cambridge, CB3 0WA, UK\\
    \ttfamily ${}^{\otimes}$ Kavli Institute for Cosmology Cambridge, \\
    \ttfamily $\phantom{{}^{\dagger}}$  Madingley Road, Cambridge, CB3 0HA, UK\\
    \ttfamily ${}^{\dagger}$ Asociaci{\'o}n Cultural La Ricotta, Espazo La Ricotta, \\
    \ttfamily $\phantom{{}^{\ddagger}}$ As Barreiras, Castro Caldelas, E-32769 Galicia-Spain\\
    \ttfamily ${}^{\ddagger}$ Instituto Galego  de F{\'i}sica de Altas Enerx{\'i}as IGFAE, \\
    \ttfamily $\phantom{{}^{\ddagger}}$ Universidade de Santiago de Compostela, E-15782 Galicia-Spain\\ 
    \ttfamily ${}^{\diamond}$ Scuola Normale Superiore, \\
    \ttfamily $\phantom{{}^{\ddagger}}$ Piazza dei Cavalieri 7, 56126, Pisa, Italy\\[0.4cm]
    \small \ttfamily andrew.mcleod@ed.ac.uk, m.capuano@herts.ac.uk, chandra@itp.ac.cn, esperc@tcd.ie, \\
      \small \ttfamily a.palazio@herts.ac.uk,  jonah.stalknecht@matfyz.cuni.cz, ysy23@cam.ac.uk, \\
      \small \ttfamily pablowellinhouse@sindominio.net, guilherme.leitepimentel@sns.it \\ [0.75cm]}

\Abstract{These lectures provide an introduction to modern techniques for studying the analytic properties of loop-level scattering amplitudes and correlation functions in perturbative quantum field theory. We begin by developing the ideas behind Landau analysis, which identifies the location and nature of kinematic singularities in Feynman integrals (the universal building blocks that contribute to perturbative amplitudes and correlators). We then show how fundamental principles such as unitarity and causality constrain the analytic structure of Feynman integrals, and develop methods for working out some of these constraints in practice. Finally, we connect these observations to the properties of iterated integrals and their symbol calculus, to show how bootstrap methods can be used to determine the functional form of Feynman integrals from just knowledge of their analytic structure. These notes are based on a series of lectures held during the \href{https://indico.sns.it/event/91/}{{\it La Ricotta} Summer School: {\it The Disordered Universe 2025}} from 21~July to 1~August~2025.}

\makepapertitle

\body

\version\versionno

\newpage

\section*{Preface}
This article is a chapter from the book {\it The Disordered Universe: Open Systems, Scattering Amplitudes, de Sitter Space, and Disordered Systems in Cosmology}~\cite{Benincasa:2027tdu}, which collects the lectures delivered at the \href{https://indico.sns.it/event/91/}{\it La Ricotta Summer School: The Disordered Universe 2025}, held from 21~July to 1~August~2025 at the \href{https://blogs.sindominio.net/laricotta/#}{\it Espazo La Ricotta} in As Barreiras (Castro Caldelas, Spain). These lecture notes are intended as a pedagogical introduction to topics that are becoming increasingly relevant in cosmology. 

\newpage

\setcounter{tocdepth}{2}
\tableofcontents

\newpage


\section{Introduction and Motivation}\label{sec:Intro}

In this set of lectures, we provide an overview of modern techniques for studying the singularity structure and discontinuities of loop-level scattering amplitudes and related quantities in perturbative quantum field theory. While this topic is best developed in the context of amplitudes, most of the techniques discussed here can also be applied to integrals that appear in other physics contexts, such as cosmology or gravitational wave physics.

\subsection{Conceptual Motivation: Physical Principles}

There are numerous reasons for us to study the analytic structure of observables in quantum field theory. Perhaps the most important reason is that these analytic properties encode many of the nontrivial implications of basic physical principles like \emph{causality}, \emph{locality}, and \emph{unitarity}. In fact, the singularities of Feynman integrals occur precisely on the support of classically-possible scattering configurations. For example, if we consider a bubble diagram with equal internal masses, some of the singularities we encounter will arise where it first becomes energetically possible to create the pair of virtual particles that travel through the loop:

\begin{figure}[H]
	\centering
	\includegraphics{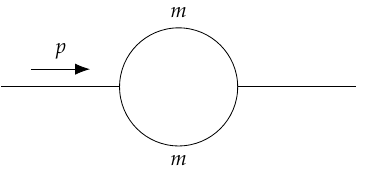}
\end{figure}

\begin{figure}[t]
\centering
\begin{subfigure}[t]{0.47\textwidth}
	\centering
	\includegraphics{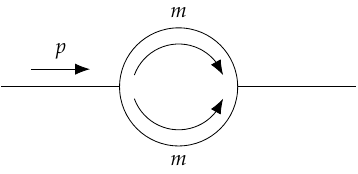}
	\caption{The threshold singularity at $p^2=(2m)^2$, which appears on the physical Riemann sheet.}
	\label{fig:threshold}
\end{subfigure}
\hfill
\begin{subfigure}[t]{0.47\textwidth}
	\centering
	\includegraphics{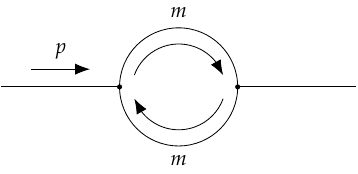}
	\caption{The pseudo-threshold singularity at $p^2=0$, which does not appear on the physical Riemann sheet.}
	\label{fig:pseudo-threshold}
\end{subfigure}
\caption{The spacetime processes associated with the threshold and pseudo-threshold singularities.}
\end{figure}

\noindent As we will later see, there are in fact two singularities that can be associated with this spacetime picture. However, only one of these singularities corresponds to a scattering configuration in which both virtual particles have positive energy (and can thus be thought of as travelling forwards in time), as depicted in Figure~\ref{fig:threshold}. This singularity, which occurs at $p^2=(2m)^2$, will appear on the first (or physical) Riemann sheet. 
The second singularity we associate with the above spacetime picture occurs at $p^2=0$, and corresponds to a configuration in which one of the virtual particles has negative energy, as depicted in Figure~\ref{fig:pseudo-threshold}. To reach this singularity, we must analytically continue the Feynman integral onto a second (or higher) Riemann sheet by varying the external kinematic variables about one (or more) of the branch points on the physical Riemann sheet.

\begin{figure}[t]
	\centering
	\includegraphics{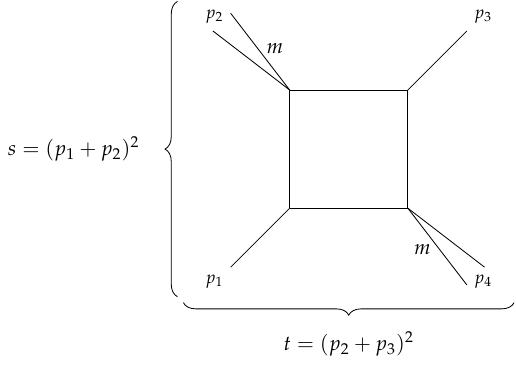}
    \caption{The box diagram $I$ with two massive legs.}
	\label{fig:box}
\end{figure}

Many of the singularities that appear in scattering amplitudes also give rise to discontinuities that are incompatible with each other. For example, consider the box diagram $I$ with two massive legs (where $p_2^2=p_4^2=m^2$ and $p_1^2=p_3^2=0$), shown in Figure~\ref{fig:box}. Kinematic constraints called the \emph{Steinmann relations} tell us that consecutive discontinuities with respect to $s$ and $t$ vanish:\footnote{Careful---this statement is no longer true when we set $m=0$~\cite{Hannesdottir:2025bss}.}
\begin{equation}
    \Disc_{t=0}\lr({\Disc_{s=0}}(I)) =\Disc_{s=0}\lr({\Disc_{t=0}}(I)) = 0\;.
\end{equation}
These constraints can be derived from causality and unitarity, and provide us with important information about the type of functions that scattering amplitudes can evaluate to. 

\subsection{Computational Motivation: Bootstrap Methods}
\label{sec:bootstrap_motivation}

In addition to the conceptual significance of understanding how physical principles are encoded in quantum field theoretic observables, knowledge of these analytic properties can provide us with powerful constraints on what functions Feynman integrals (and by extension, amplitudes and correlators) can evaluate to. This can prove to be extremely useful, since the integrals that appear in physics are often too difficult to evaluate directly. However, if we have sufficiently detailed knowledge of the analytic properties of an integral or a physical quantity of interest, we may be able to \emph{bootstrap} its functional form directly. 

This strategy has proven remarkably successful for computing (certain) perturbative amplitudes in planar maximally supersymmetric ($\mathcal{N}=4$) Yang-Mills theory. For example, six-particle amplitudes and three-particle form factors (the latter of which can be thought of as four-particle amplitudes) have been computed via these methods through 8 loops~\cite{Dixon:2022rse,Dixon:2023kop}:

\begin{figure}[h!]
	\centering
	\includegraphics{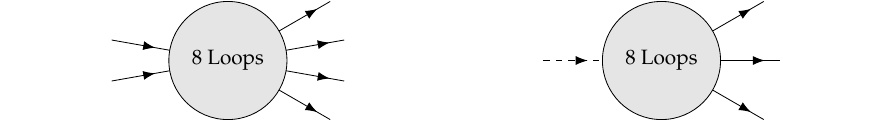}
\end{figure}

\noindent These bootstrap calculations leverage knowledge of (or educated guesses for):
\begin{itemize}
    \item the types of functions the amplitudes evaluate to, 
    \item symmetries, such as Bose symmetry and spacetime parity, 
    \item an understanding of their analytic and number-theoretic properties, and
    \item concrete calculations of their kinematic limits.
\end{itemize}
Once computed, these amplitudes have also served as a rich data set, whose study has led to the discovery of new symmetries, novel analytic constraints, and surprising dualities. It is hoped that the same type of bootstrap methods may also find similar levels of success in other areas of quantum field theory, such as in cosmology.
  
\subsection{Goal of these Lecture Notes}
The goal of these lectures is to introduce a set of techniques that have proven useful in the study of amplitudes, and that are also expected to be useful in other contexts. Thus, although the main focus will be on scattering amplitudes and Feynman diagrams, we will illustrate how these methods also generalize to other classes of integrals by drawing examples from cosmology.

An outline of the topics covered in these lecture notes is as follows:
\begin{itemize}
    \item Lecture~\hyperref[sec:sing]{1}: (Kinematic) Singularities of integrals
    \item Lecture~\hyperref[sec:disc]{2}: Discontinuities of integrals
    \item Lecture~\hyperref[sec:seqdisc]{3}: Discontinuities of discontinuities
    \item Lecture~\hyperref[sec:special]{4}: Special functions
    \item Lecture~\hyperref[sec:bootstrap]{5}: Bootstrap methods
\end{itemize}
Additionally, \hyperref[app:exer]{exercises} will be provided at the end, with worked \hyperref[app:soln]{solutions}.


\section{Singularities of Integrals}\label{sec:sing}

Making predictions in physics often involves evaluating integrals over rational functions. A prominent class of examples is provided by Feynman integrals, which appear in the study of perturbative quantum field theory. Due to the ubiquity of these types of integrals, it can be interesting and useful to diagnose where they become singular and have the potential to develop branch cuts.

\subsection{Pinches and Endpoints}
\label{sec:pinches_endpoints}

\begin{figure}[t]
	\centering
	\includegraphics{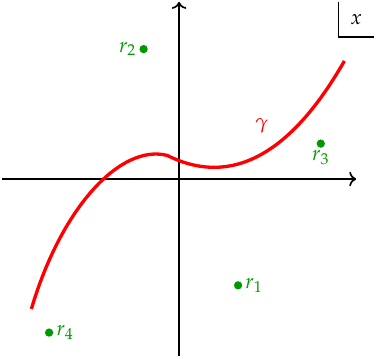}
    \caption{A possible configuration of the contour $\gamma$ and the roots $r_i(z)$ in the complex $x$-plane.}
	\label{fig:sing0}
\end{figure}

We start our investigation by considering a generic function of the form
\begin{align} \label{eq:rational_int}
    f(z) = \int_{\gamma} \frac{g(x, z)}{h(x, z)}\; \dd x\;,
\end{align}
where $g(x, z)$ and $h(x, z)$ are polynomials and $\gamma$ is an integration contour that does not depend explicitly on $z$. We assume that $f(z)$ is well-defined and finite for some value of $z$, and will be interested in understanding when $f(z)$ can become singular as we vary $z$. Since the contour $\gamma$ does not depend on $z$, this can only happen if the integrand becomes singular along this contour. These singularities occur at the roots $r_i(z)$ of $h(x, z)$, namely 
\begin{align}
    h(x,z) = (x-r_1(z))(x-r_2(z))\ldots (x-r_n(z))\;,
\end{align}
where we assume that $h(x,z)$ is a polynomial in $x$ of degree $n$. An example configuration of the integration contour $\gamma$ and the roots $r_i(z)$ in the integration plane is shown in Figure \ref{fig:sing0}.

\begin{figure}[t]
	\centering
	\includegraphics{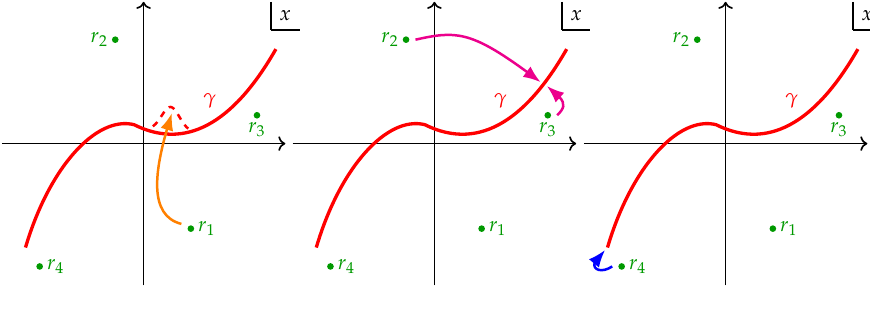}
    \caption{Ways in which the denominator roots $r_i(z)$ can intersect the integration contour as we vary the external variable $z$. \textit{Left:} The contour can be locally deformed away from the integrand-level singularity, so the integral does not become singular. \textit{Middle:} The contour is pinched between two roots in such a way that it cannot be deformed out of the way; this results in a pinch singularity. \textit{Right:} One of the roots intersects an endpoint of the integration contour; the contour cannot be deformed away from the endpoint, so this leads to an endpoint singularity.}
	\label{fig:sings}
\end{figure}

As we vary $z$, the roots $r_i(z)$ will move around in the $x$ plane. When the roots intersect the contour $\gamma$, one of three things can happen:
\begin{enumerate}
    \item[(i)] One of the roots crosses the contour. In this case, the contour can be deformed away and $f(z)$ will not become singular.

    \item[(ii)] Two or more roots can approach the contour from opposite sides. In this case, the contour cannot be deformed away, and the integral inherits the singularity of the integrand. This is a \emph{pinch singularity}.

    \item[(iii)] One or more roots can intersect one of the endpoints of $\gamma$. Again, the contour cannot be deformed away, and the integral inherits the singularity of the integrand. This is an \emph{endpoint singularity}. 

\end{enumerate}
These different situations are illustrated in Figure \ref{fig:sings}.

Thinking about the pinch and endpoint configurations described above, we can formulate a set of necessary conditions for $f(z)$ to become singular:
\begin{gather} 
h(x,z)=0 \label{eq:vanishing_denom} \\
(x-\g(0))(x-\g(1)) \lr({\frac{\ptl}{\ptl x} h(x, z)}) = 0\,, \label{eq:pinch_endpoint}
\end{gather}
where $\gamma(0)$ and $\gamma(1)$ represent the endpoints of the contour (if one or both of these endpoints doesn't exist, we simply drop the corresponding factor(s)). The factors
\begin{equation*}
	(x-\g(0))(x-\g(1))
\end{equation*}
in the second equation denote endpoint conditions, while the factor 
\begin{equation*}
	\lr({\frac{\ptl}{\ptl x} h(x, z)})
\end{equation*}
encodes the pinch condition. To see why the vanishing of the factor $\ptl_x h(x,z)$ allows us to probe for pinch singularities, we note that this factor will vanish in the $x$ plane wherever two or more roots coincide, since
\begin{align}
    \frac{\ptl}{\ptl x}h(x,z) = \sum_{i=1}^n \frac{h(x,z)}{x-r_i(z)}\;,
\end{align}
where each term on the right involves all but one root. Equation~\eqref{eq:pinch_endpoint} thus just constitutes a \textit{necessary} condition for $f(z)$ to become singular. It is not a sufficient condition, since we have not incorporated any information about $g(x, z)$ or about the homotopy class of $\g$.

\subsection{Example 1: A Onefold Integral}
\begin{figure}[t]
	\centering
	\includegraphics{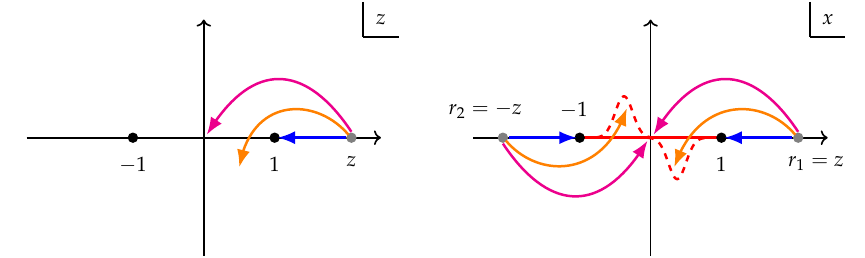}
    \caption{\textit{Left:} Paths along which we can analytically continue $f(z)$ in the $z$ plane. \textit{Right:} We can diagnose whether the integral $f(z)$ becomes singular along each analytic continuation path by following how the roots $r_1$ and $r_2$ intersect the integration contour. }
	\label{fig:ex1}
\end{figure}

Let us see this in a simple example. Consider the integral
\begin{align}\label{eq:example1}
    f(z) = \int_{-1}^1 \frac{\dd x}{(x+z)(x-z)}\;.
\end{align}
As shown in Figure \ref{fig:ex1}, the roots that appear in the denominator occur at $r_1(z)=z$ and $r_2(z)=-z$, and move accordingly in the $x$ plane as the external parameter $z$ is varied. In particular, if we vary $z$ such that it passes through the line segment between -1 and 1, the roots will intersect the integration contour. Depending on where these roots pass through the real axis, we see different behavior:
\begin{enumerate}
    \item[(i)] If $z$ crosses the real axis away from the points $\{-1,0,1\}$, the contour can be deformed away, and hence the integral is not singular.
    \item[(ii)] As $z$ approaches $0$, the two roots approach the integration contour from opposite sides, leading to a pinch singularity.
    \item[(iii)] If $z$ approaches $1$ or $-1$, both roots intersect the endpoints of the integration contour. This leads to a pair of endpoint singularities.
\end{enumerate}

\noindent Indeed, one can check that this integral evaluates to
\begin{align}
    f(z) = \frac{\log(1-z^{-1})- \log(1+z^{-1})}{z}\;,
\end{align}
which has logarithmic singularities at $z=\pm 1$ and a simple pole at $z=0$ (where the numerator evaluates to $\ci \pi$).

\subsection{Example 2: A Twofold Integral}
\begin{figure}[t]
	\centering
	\includegraphics{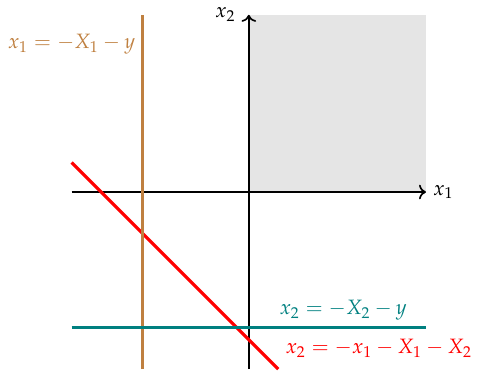}
    \caption{The integration region (shaded) and singular loci that appear in the integral \eqref{eq:cosmo-corr-ex}, for positive values of $X_1$, $X_2$, and $y$.}
	\label{fig:int-reg}
\end{figure}

We will now consider a higher-dimensional example that appears in cosmology. When evaluating the trispectrum of a conformal scalar in de Sitter \cite{Arkani-Hamed:2017fdk-ch03}, one encounters an integral of the form
\begin{align}\label{eq:cosmo-corr-ex}
    \int_{x_1,x_2\geq 0}\frac{\dd x_1\dd x_2}{(x_1+X_1+x_2+X_2) (x_1+X_1+y) (x_2+X_2 + y)} \equiv \int_{x_1,x_2\geq 0 }\frac{\dd x_1\dd x_2}{M}\;,
\end{align}
where $X_1$, $X_2$, and $y$ take positive values in physical kinematics. A sketch of the integration region and the loci where the integrand  becomes singular (namely, where the denominator $M$ vanishes), is depicted in Figure \ref{fig:int-reg}.

Where can singularities occur for this integral in the space of kinematic variables $X_1$, $X_2$, and $y$? The most obvious candidates are endpoint singularities where the surface defined by the vanishing of one of the denominator factors intersects the boundary of the integration contour. For instance, the $x_1 + X_1 + x_2 + X_2 = 0$ surface seems to intersect the integration boundary twice when we analytically continue to a kinematic region in which $X_1+X_2< 0$: first when $x_1=0$ and $x_2 = -X_1-X_2$, and again when $x_2 = 0$ and $x_1 = -X_1-X_2$. Are these valid endpoint singularities? Actually, they are not. To see this we need to think about what freedom we still have to deform our integration contour out of the way. 

First, let us split up each of the $x_i=0$ conditions that define the integration endpoints in~\eqref{eq:cosmo-corr-ex} into their real and imaginary parts: $\text{Re}(x_i) = 0$ and $\text{Im}(x_i) = 0$. Similarly, we can break up the constraint  $x_1 + X_1 + x_2 + X_2 = 0$ that defines our singular surface into its real and imaginary parts. Restricting our attention to kinematics in which $X_1 + X_2$ is real and negative (which is where we would expect an endpoint singularity to occur, since the initial contour ranges over real values of $x_1$ and $x_2$), we have that the $x_1 + X_1 + x_2 + X_2 = 0$ surface satisfies $\text{Im}(x_1) + \text{Im}(x_2) = 0$. Let's now think about how this surface might intersect the $x_1=0$ integration endpoint, which we worried would happen when $x_2 = -X_1 -X_2$. Recalling that the imaginary part of this endpoint satisfies $\text{Im}(x_1)=0$, we see that $x_1 + X_1 + x_2 + X_2 = 0$ only passes through $x_1=0$ when $-\text{Im}(x_2) = \text{Im}(x_1) = 0$ for real $X_1 + X_2<0$. However, since we are not near the $x_2=0$ integration endpoint, there is no condition on where in the complex $x_2$ plane the integration contour must be; in particular, the integration contour need not satisfy the condition $\text{Im}(x_2)=0$ when $\text{Re}(x_2)>0$. Thus, we are free to locally deform the integration contour in the $\text{Im}(x_2)$ direction, such that it satisfies $\text{Im}(x_2)\neq0$ even while $\text{Im}(x_1) = 0$. In this way, we manage to deform the integration contour away from the $x_1 + X_1 + x_2 + X_2 = 0$ surface, since the latter only intersects the $x_1=0$ surface when $\text{Im}(x_2) = \text{Im}(x_1) = 0$. In the same way, we can avoid an endpoint singularity at $x_2 = 0$ and $x_1 = -X_1-X_2$ by locally deforming the contour in the $\text{Im}(x_1)$ direction.\footnote{It is a worthwhile exercise to rerun this argument for complex values of $X_1$ and $X_2$, to convince yourself that no endpoint singularity develops even when these variables are allowed to have nonzero imaginary parts.}

To rephrase this in a more active language, let's describe how these candidate endpoint singularities can be avoided if we analytically continue the $x_1 + X_1 + x_2 + X_2 = 0$ surface to intersect the positive real quadrant in Figure~\ref{fig:int-reg}. If we start in the physical region, $X_1$ and $X_2$ both take positive, real values. Keeping $X_2$ fixed, we can vary $X_1$ by first giving it a small positive imaginary part; we then decrease the real part of $X_1$ until $\text{Re}(X_1) + X_2 < 0$. Finally, holding the real part of $X_1$ fixed, we decrease its imaginary part until $X_1$ is purely real again. At this point, the singular surface $x_1 + X_1 + x_2 + X_2 = 0$ intersects the undeformed integration contour. However, as discussed above, by deforming the integration contour in the negative imaginary $x_2$ direction near the point $x_1=0, \ x_2 = -X_1-X_2$ and in the negative imaginary $x_1$ direction near the point $x_2=0, \ x_1 = -X_1-X_2$, these two surfaces can be made to never actually intersect. 

Of course, there's precisely one point where this deformation of the integration contour cannot be carried out. At the intersection of the $x_1$ and $x_2$ endpoints, the contour is required to satisfy $\text{Re}(x_1) = \text{Im}(x_1) = \text{Re}(x_2) =\text{Im}(x_2) = 0$. Thus, while no endpoint singularity arises for $X_1+X_2< 0$, an endpoint singularity does arise at $x_1=x_2=0$ when $X_1+X_2=0$. 

We can carry out a similar argument to convince ourselves that the other two denominator factors give rise to endpoint configurations. Thus, there exist three endpoint singularities
\begin{enumerate}
    \item[(1)] $X_1+X_2 = 0$: endpoint singularity at $x_1=x_2 = 0$
    \item[(2)] $X_1+y = 0$: endpoint singularity at $x_1= 0$
    \item[(3)] $X_2+y = 0$: endpoint singularity at $x_2= 0$
\end{enumerate}
that arise just from the intersection of one of the denominator factors with the integration endpoints.

There also exist singularities that arise from two or more of the denominator factors pinching the integration contour. For example, where the vanishing locus of the denominator factors $(x_1+X_1+x_2+X_2)$ and $(x_2+X_2 + y)$ intersect, we have that
\begin{align}
    x_2 = -X_2-y = -x_1-X_1-X_2 \implies x_1= y -X_1\;.
\end{align}
Since we have the freedom to give $-X_2-y$ and $-x_1-X_1-X_2$ small imaginary parts of different sign, we can indeed arrange a situation in which these surfaces approach the integration contour from opposite sides. However, when these surfaces are coincident, they intersect where $\text{Im}(x_1) = \text{Im}(y-X_1)$. Thus, we can avoid this pinch by locally deforming the integration contour in $\text{Im}(x_1)$ direction, so that $\text{Im}(x_1) \neq \text{Im}(y-X_1)$. That is, we can do this everywhere \emph{except} at $x_1=0$ (since here, the contour is restricted to satisfy $\text{Im}(x_1) = 0$). Thus, we see that a singularity will arise when $0 = x_1 = y-X_1$. By symmetry, we will also get a singularity due to a pinch between the $(x_1+X_1+x_2+X_2)$ and $(x_1+X_1+y)$ surfaces:
\begin{enumerate}
    \item[(4)] $X_1-y = 0$: pinch singularity where $x_2 = -X_2-y$ and $x_1= 0$
    \item[(5)] $X_2-y = 0$: pinch singularity where $x_1 = -X_1-y$ and $x_2= 0$
\end{enumerate}
Conversely, the pair of surfaces $x_1+X_1+y=0$ and $x_2+X_2 + y=0$ cannot pinch the integration contour, because it is sufficient to deform the integration contour such that either $\text{Im}(x_1) \neq \text{Im}(-X_1-y)$ or $\text{Im}(x_2) \neq \text{Im}(-X_2-y)$ to avoid this pinch. The only place where neither of these deformations can be carried out is when we are at the combined $x_1=x_2=0$ endpoint, but at this point we are just sitting on top of the pair of endpoint singularities (1) and (2) considered above. 

Finally, there is a singularity involving all three denominator factors
\begin{align}
    -X_1-X_2 = x_1+x_2 = -X_1-y -X_2-y \implies -2y=0\;,
\end{align}
giving us
\begin{enumerate}
    \item[(6)] $y = 0$: pinch singularity where $x_1= -X_1$ and $x_2 = -X_2$
\end{enumerate}
We leave it as an exercise to the reader to convince themselves that this is a genuine pinch, namely that the contour cannot be deformed out of the way.

Carrying out this type of geometric reasoning in higher dimensions becomes increasingly difficult. Thus, in practice, it is easier to work with algebraic conditions. One can verify that all of the singular configurations above can all be reproduced if we simply require the pair of constraints~\eqref{eq:vanishing_denom} and~\eqref{eq:pinch_endpoint} we derived in the one-variable case, for both integration variables. Namely, we solve the equations
\begin{align}
    M =0 \;, \qquad x_1\lr({\frac{\ptl}{\ptl x_1} M})  =0 \;, \qquad x_2\lr({\frac{\ptl}{\ptl x_2} M})  =0 \;,
\end{align}
which require that there is an endpoint or a pinch singularity in both integration variables simultaneously. 

\subsection{The Landau Equations}
We will now use these ideas to derive the \textit{Landau equations}, which specifically describe where Feynman integrals can become singular. A Feynman integral is an integral that takes the form
\begin{align}
    I_G(p_i,m_e) = \int_{-\infty}^\infty \frac{\dd^D\ell_1\ldots\dd^D\ell_L}{\prod_{e=1}^E (q_e^2-m_e^2+i\varepsilon)}\;,
    \label{eq:Feynman_int}
\end{align}
where $G$ is a Feynman graph with $L$ loops and $E$ internal edges, $D$ is the spacetime dimension, and $m_e$ and $q_e$ are the mass and momentum flowing through edge $e$. The momenta $q_e$ depend linearly on the loop momenta $\ell_i$ and the external momenta $p_i$.

To establish the standard form of the Landau equations, we first reformulate the above integral by introducing Feynman parameters $\a_i$,
\begin{align}\label{eq: Feynman parametrization}
     I_G(p_i,m_e) \sim  \int_{-\infty}^\infty \dd^D\ell_1\ldots\dd^D\ell_L \int_{0}^\infty \dd\a_1\ldots \dd \a_E\frac{\de(1-\sum \a_e)}{M^E}\;,
\end{align}
where $M=\sum_{e=1}^E \a_e (q_e^2-m_e^2+\ci\varepsilon)$. To find the singularities of $I_G(p_i,m_e)$, we first require that each Feynman parameter is either in an endpoint or pinch configuration. This gives us the set of constraints
\begin{align}
    \a_e \lr({\frac{\ptl}{\ptl \a_e} M}) = \a_e (q_e^2-m_e^2+i\varepsilon) = 0 \;, 
\end{align}
which we impose for all integration variables $\alpha_e$ simultaneously. Jointly, these conditions imply $M = 0 $, thus the constraint~\eqref{eq:vanishing_denom} is automatically satisfied. Next, we require each loop momentum to similarly be in an endpoint or pinch configuration. Since the integrals over loop momenta have no endpoints, we simply have
\begin{align}
    \frac{\ptl}{\ptl \ell_i^\mu} M  = \sum_{e\in{\rm loop}} \pm\a_eq_e^\mu = 0\;,
\end{align}
where the sum is over all edges a given loop momentum $\ell_i$ flows through, and the signs depend on the relative orientation of $\ell_i^\mu$ and $q_e^\mu$. Together, these constitute the Landau equations~\cite{nakanishi1959,Landau:1959fi,Bjorken:1959fd}:
\begin{shade}
\begin{align}
    \label{eq:landua1}\a_e(q_e^2-m_e^2+i\varepsilon) &= 0 \qquad \text{for all edges $e$,} \,\\
    \label{eq:landua2}\sum_{e\in{\rm loop}} \pm\a_eq_e^\mu &= 0  \qquad \text{for every independent loop in $G$.}    
\end{align}
\end{shade}
\noindent Formally, these equations define a variety on which our original Feynman integral can become singular. Note that this variety lives in the combined space of integration variables and external kinematic variables. Sometimes, it is useful to talk about where a Feynman integral becomes singular just in the space of external kinematic variables, in which case we are effectively saying ``at this kinematic point, there exists some locus in the space of integration variables where a singularity arises.''  In other situations, especially when considering the sequential discontinuities of Feynman integrals, it will be useful to keep track of exactly where in the space of integration variables each singularity arises.

\begin{figure}[t]
	\centering
	\includegraphics{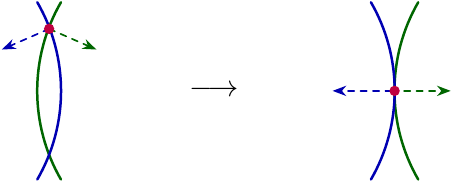}
    \caption{\textit{Left:} A situation in which the on-shell surfaces associated with a pair of propagators, depicted in blue and green, intersect in a generic way. In this configuration, there exists no linear relation between the normals to these surfaces at the points of intersection, and correspondingly the integration contour can be deformed to avoid getting pinched by these intersections. \textit{Right:} A situation in which the on-shell surfaces intersect tangentially, so there exists a linear relation between their normals at the intersection point. If the contour is caught between these surfaces, it cannot be deformed away, and the integral will become singular.}
	\label{fig:normals}
\end{figure}

Let us next pause for a moment to ask why it is that the Landau equations, which describe the singularities of the original loop-momentum integral, depend on variables $\a_e$ that appear nowhere in~\eqref{eq:Feynman_int}. It turns out these variables take on a geometric meaning when we think about the original loop momentum integral. Namely, the variable $\a_e$  parametrizes whether or not the on-shell surface $q_e^2 - m_e^2 + i \epsilon = 0$ is involved in pinching the integration contour. In particular, when a given solution to the Landau equations requires $\a_e \neq 0$, we know from~\eqref{eq:landua2} that we are considering a singularity that requires $q_e^2 - m_e^2 + i \epsilon = 0$. Moreover, equation~\eqref{eq:landua2} tells us that there exists a linear relation between the vectors normal to the on-shell surfaces that are involved in the pinch, which is necessary for these on-shell surfaces to have the possibility of trapping the contour.\footnote{In more detail, we have that the total differential of each on-shell surface is $d(q_e^2-m_e^2+i\varepsilon)=2 q_e \cdot d q_e = 2\sum_i \pm q_e \cdot d \ell_i= 0$, where the signs again depend on the relative orientation of each $\ell_i^\mu$ and $q_e^\mu$. Thus, the coefficients $\pm q_e^\mu$ define the normal to the on-shell surface $q_e^2-m_e^2+i\varepsilon=0$, so Equation~\eqref{eq:landua2} is just telling us that there exists a linear relation among these normal vectors. Stated differently, these on-shell hypersurfaces do not intersect transversely.} See Figure \ref{fig:normals} for a simple illustration of what this looks like geometrically. Conversely, when $\a_e=0$, the corresponding propagator need not be on shell, and the corresponding momentum $q_e^\mu$ drops out of the constraint~\eqref{eq:landua2}. 

Finally, looking forward, let us highlight that there are specific labels that are often used to describe different types of solutions to the Landau equations:
\begin{itemize}
    \item \emph{second type} singularities refer to singularities that occur at infinite values of the loop momenta
    \item \emph{$\a$-positive} singularities are those that occur for non-negative, real values of all Feynman parameters.
\end{itemize}
As we will see, $\a$-positive singularities are the most physically-meaningful solutions to the Landau equations, while the interpretation of second type singularities (and more generally, singularities that occur for complex values of the Feynman parameters) is less direct.

\subsection{Coleman-Norton Interpretation}
For solutions to the Landau equations involving only massive internal particles, a physical interpretation was offered by Coleman and Norton~\cite{Coleman:1965xm}. Starting from the appropriate Feynman diagram $G$ and a solution to the Landau equations, we can contract all edges that are not on shell. The resulting contracted diagram $G'$ is called a {\it Landau Diagram}, and can be interpreted as a classically allowed scattering process. 
Namely, every loop represents a closed loop in spacetime, with the identification
\begin{align}
    \a_e= \frac{\tau_e}{m_e}\;,
\end{align}
where $\tau_e$ is the proper time traveled by the virtual particle represented by edge $e$. Note that the momenta in these loops may not live in Minkowski signature, therefore we have a generalized notion of ``classically allowed'' in mind. In subsequent sections, we will use Landau diagrams to represent which propagators are on shell for a given singularity (or solution to the Landau equations). In general, there will be many singularities we can associate with each Landau diagram.

\subsection{``Landau Equations'' for Cosmological Correlators}

Just to illustrate that the ideas we are developing here are general, let us also formulate a set of ``Landau equations'' for (the site energy integrals of) cosmological correlators, which take the form
\begin{align} \label{eq:cosmological_correlators}
    \int_{x_i\geq 0 } \frac{\dd x_1\ldots \dd x_n}{S_1\ldots S_m}\;,
\end{align}
where the $S_j$'s depend linearly on the integration variables $x_i$ as well as on external kinematic variables. An example of such an integral was given in~\eqref{eq:cosmo-corr-ex}.

By analogy with the case of scattering amplitudes, we could introduce Feynman parameters, but there is no strict need. Without them, we already have
\begin{shade}
    \begin{align}
    \prod_{j=1}^m S_j &= 0\;,\\
    x_i \frac{\ptl}{\ptl x_i}\lr({\prod_{j=1}^m S_j}) &= 0\;,
\end{align}
\end{shade}
\noindent where the last equation is required simultaneously for each integration variable $x_i$. Note that the first equation does not require all denominator factors to vanish (just their product), while the second equation encodes the conditions for either an endpoint or pinch singularity in each integration variable. Further examples of deriving and solving the Landau equations in different integral representations will be presented in the tutorial problems.

\subsection{Estimating the Nature of Singularities}
Let us also briefly highlight that, in addition to writing down equations~\eqref{eq:landua1} and~\eqref{eq:landua2}, Landau outlined a method for predicting the \emph{nature} of the singularities that appear in Feynman integrals~\cite{Landau:1959fi}. We will work through an example in which we apply this method in the first tutorial (\ref{sec: tutorial 1}); for now, we just quote the general result from Landau's paper, which applies to Feynman integrals $I_G$ that involve only generic masses (meaning all particles are massive, and the mass associated with each line in our diagram is distinct). Namely, if $I_G$ becomes singular as some polynomial of kinematic variables $\varphi \to 0$, we can predict the leading non-analytic behavior exhibited by the integral to be
\begin{equation} \label{eq:singularity_nature}
I_G(\varphi\rightarrow0)\propto 
\begin{cases}
\varphi^{\gamma}\log\varphi & \text{if $\gamma\in \mathbb{Z}$ and $\gamma\geq0$} \\
\varphi^{\gamma} & \text{otherwise}
\end{cases}\,,
\end{equation}
where $\gamma=\frac{1}{2}(LD-E-1)$, $D$ is the number of spacetime dimensions, and $L$ and $E$ denote the number of loops and internal propagators that are left after all the propagators that are not required to be on shell have been contracted (which is to say, the number of loops and edges in the corresponding Landau diagram).

For example, for the threshold and pseudo-threshold singularities represented by the diagrams in Figures~\ref{fig:threshold} and~\ref{fig:pseudo-threshold}, we have that $L=1$ and $E=2$. So, in $D$ spacetime dimensions, the bubble integral is predicted to behave as
\begin{equation}\label{eq:bubble-singularities}
I_G(\varphi\rightarrow0)\propto 
\begin{cases}
\varphi^{D-3}\sqrt{\varphi}\log \varphi, & \text{when $D-3\geq0$ even} \\
\varphi^{D-3}\sqrt{\varphi}, & \text{otherwise.}
\end{cases}\quad
\end{equation} 
We will see later that this is indeed true. Note that the bubble diagram also has singularities that only involve a single on-shell surface; for these singularities we have $L=1$ and $E=1$.

We emphasize that the result~\eqref{eq:singularity_nature} only holds when all of the edges that appear in $G$ have distinct, nonzero masses. This is simply because different singularity surfaces start to intersect when we set these masses equal to each other, or even to zero, which makes the analysis more difficult. For more on this topic, we refer the reader to~\cite{pham2011singularities,Hannesdottir:2021kpd,Hannesdottir:2024hke,Vergu:2025mag}.


\section{Discontinuities of Integrals}\label{sec:disc}
In this lecture we describe how to think about and compute the discontinuities of Feynman integrals. While it is common to introduce discontinuities as the difference between the values a function takes on different sides of its branch cuts, we will here generalize this notion by defining discontinuities in terms of analytic continuation. This will ensure that discontinuities remain well-defined on the full complex plane (not just on branch cuts), which will prove necessary when we start considering discontinuities of discontinuities in the next lecture. We will also see that only those solutions to the Landau equations that are $\a$-positive are expected to appear on the first (physical) Riemann sheet.
 
\subsection{Multivalued Functions }
Unitarity tells us that scattering amplitudes, and related quantities such as correlators, are multi-valued functions.
We can show this by considering a generic unitary S-matrix. Namely, separating out the interacting part of the S-matrix and requiring it to be unitary, we find
\begin{equation}
    S = \id + i T\,,
\end{equation}
\begin{equation}
    S S^\dagger = \id \implies \frac{1}{2i}( T - T^\dagger) = \frac{1}{2}T T^\dagger\,.
\end{equation}
In the forward-scattering limit, where the incoming and outgoing states are identical, we therefore have
\begin{equation}
\Im(T_{in \to in}) = \frac{(2\pi)^D}{2} \sum \hspace{-0.5cm}\int \abs{T_{in \to I}}^2 \de^D(p_{in}-p_I)\,,
\label{eq:optical_thm}
\end{equation} 
where the sum/integral is over all possible intermediate states $I$. Now let $s_0$ denote the energy required to produce the lightest multi-particle intermediate state. For $s<s_0$, the right side of \eqref{eq:optical_thm} must vanish due to the conservation of energy, as imposed by the Dirac delta function. Hence $T_{in \to in}(s)$ is manifestly real for $s < s_0$, and we can use the Schwarz reflection principle to analytically continue $T_{in \to in}(s)$ to the whole complex plane using the relation
\begin{equation}
(T_{in \to in} (s^*) )^* = T_{in \to in}(s)\,, \quad s< s_0 \,. 
\label{eq:schwarz}
\end{equation}
Now, let us analytically continue \eqref{eq:schwarz} to $s>s_0$. Since we know that $T_{in \to in}(s)$ is no longer real for $s > s_0$, we can use \eqref{eq:schwarz} to relate $T_{in \to in}(s)$ above and below the real line as follows,
\begin{align}
\Im(T_{in \to in}(s + i \varepsilon)) = & \frac{1}{2i} ( T_{in \to in}(s+ i \varepsilon) - T^*_{in \to in}(s+i \varepsilon) )\\
= & \frac{1}{2i} ( T^*_{in \to in}(s-i \varepsilon) - T_{in \to in}(s-i \varepsilon) ) \\
= & - \Im( T_{in \to in}(s-i \varepsilon) )\,,
\end{align}
while
\begin{equation}
\Re( T_{in \to in}(s + i \varepsilon) ) = \Re(T_{in \to in} ( s-i \varepsilon))\,.
\end{equation}
Therefore, we see that there exists a nonzero discontinuity $\Disc_{s \to s_0} {(T_{in \to in} (s \geq s_0) )} = 2 i \Im(T_{in \to in} (s+i \varepsilon) )$ starting at $s = s_0$.

\begin{figure}[t]
	\centering
	\includegraphics{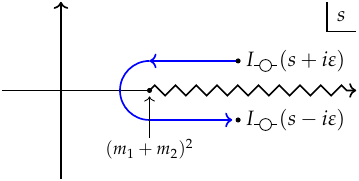}
    \caption{Complex plot of the massive bubble integral, highlighting the branch structure near the threshold singularity. The path along which $I_{\bubbletry}(s+i\varepsilon)$
    is analytically continued to $I_{\bubbletry}(s-i\varepsilon)$ 
    is shown in blue. While the location of the branch point is unambiguous, the placement of the branch cut is just a convention.}
	\label{fig:bubble_branch}
\end{figure}

Let us illustrate how we can compute such discontinuities at fixed perturbative order, using the example of the massive bubble integral in $D=2$:
\begin{equation}
\begin{tikzpicture}[baseline]
    \draw (-3, 0) -- (-1, 0);
    \node at (-2, 0.5) {$s = p^2$};
    \draw (1, 0) -- (3, 0);
    \draw (0,0) circle (1);
    \node at (0,1.25) {$m_1$};
    \node at (0, -1.25) {$m_2$};
    \filldraw (-1, 0) circle (1pt);
    \filldraw (1, 0) circle (1pt);
\end{tikzpicture}
\end{equation}
The threshold singularity for this integral occurs when $p^2 = s = \left(m_1 + m_2\right)^2$, and gives rise to the branch structure shown in Figure \ref{fig:bubble_branch}. One way to compute the discontinuity with respect to $s = \left(m_1 + m_2\right)^2$ is by directly evaluating the integral on either side of the branch cut, namely
\begin{equation}
    \Disc_{s- \left(m_1 + m_2\right)^2} I_{\bubble}  = I_{\bubble}\left(s-i \varepsilon\right) - I_{\bubble}\left(s+i \varepsilon\right) \,.
\end{equation}
Alternatively we can analytically continue $I_{\bubble}(s+i\varepsilon )$ around the branch point at $s = ( m_1 + m_2)^2$, as shown in blue in Figure \ref{fig:bubble_branch}, and take the difference between the value of our integral before and after carrying out this analytic continuation:
\begin{equation}
\Disc_{s- \left(m_1 + m_2\right)^2} I_{\bubble} = \Big( \begin{tikzpicture}[scale=1.2, baseline={([yshift=-.5ex]current bounding box.center)}]
    \filldraw (0,0) circle (1pt);
    \draw[->] (0.173,0.1) arc (30:330:0.2);
\end{tikzpicture} - \mathbb{1} \Big) I_{\bubble}\,.
\end{equation}The advantage of this second method is that it is well-defined for any value of $s$; we can analytically continue $s$ around the branch point at $s = ( m_1 + m_2)^2$, no matter what the initial value of $s$ is in the complex plane. This will prove important in subsequent lectures, where we will want to make sense of sequences of discontinuities such as
\begin{equation}
\Disc_s \Disc_{s-4m^2} \Disc_m^2 (I)
\end{equation}
for arbitrary values of $s$ and $m^2$.

\subsection{Bubble Example}
\label{sec:massive_bubble}

Let's see what this looks like more concretely, in the case of the bubble. Starting from the mixed representation schematically outlined in~\eqref{eq: Feynman parametrization}, we have
\begin{equation}
    I^{D=2}_{\bubble} = \int_{-\infty}^\infty \dd^D \ell \int_{0}^\infty \dd\a_1 \int_0^\infty \dd \a_2 \frac{\de(1-\a_1 - \a_2)}{\left(\a_1(q_1^2-m_1^2+i\varepsilon) + \a_2(q_2^2-m_2^2+i\varepsilon)\right)^2}\;.
    \label{eq:bubble_feynman}
\end{equation}
Integrating over the loop momentum and $\alpha_2$ (the latter of which simply gets rid of the delta function), we arrive at the onefold integral
\begin{equation}
I^{D=2}_{\bubble} = \int_0^1 \frac{\dd\a}{F} \,, 
\label{eq:bubble_2d_int}
\end{equation}
where 
\begin{equation}
    F = s \a (1-\a) -m_1^2 \a - m_2^2(1-\a) + i \varepsilon\,
\end{equation}
is sometimes called the second Symanzik polynomial. Note that we have dropped the index on the remaining integration variable $\alpha_1$.

In this representation, the Landau equations are given by the conditions we derived in section~\eqref{sec:pinches_endpoints}; concretely, these correspond to
\begin{equation}
F = 0\,, \quad \alpha (1-\alpha)\frac{\partial}{\partial \alpha}F=0\,.
\end{equation}
For now, let us focus on singularities that arise from pinches of the $\alpha$ integration contour, which means focusing on the solutions that set $\frac{\partial}{\partial \alpha}F=0$. To find these singularities, we first identify the two roots of $F$ that solve the first equation:
\begin{equation}
\a_{\pm} = \frac{s + m_2^2 - m_1^2 \pm \sqrt{s^2 - 2 s (m_1^2+m_2^2) + (m_1^2-m_2^2)^2 + i s \,\varepsilon}}{2s}\,.
\label{eq:bubble_2d_roots}
\end{equation}
If we then impose the vanishing of the derivative of $F$, we find two solutions: the threshold singularity
\begin{equation}
s =  (m_1 + m_2)^2 - i \varepsilon\,, \quad  \a_\pm =  \frac{m_2}{m_2 + m_1} + i \varepsilon \,\sgn{m_2-m_1}\,,
\label{eq:bubble_2d_thresh}
\end{equation}
and the pseudo-threshold singularity
\begin{equation}
s =  (m_1-m_2)^2 + i \varepsilon\,, \quad  \a_{\pm} =  \frac{m_2}{m_2-m_1} - i \varepsilon \,\sgn{m_2-m_1}\,.
\label{eq:bubble_2d_pseudo}
\end{equation}
Geometrically, as $s$ approaches the values written in either \eqref{eq:bubble_2d_thresh} or \eqref{eq:bubble_2d_pseudo}, the two roots \eqref{eq:bubble_2d_roots} coincide at $\frac{m_2}{m_2\pm m_1}$, and hence these roots may pinch the integration contour. Note that the lack of symmetry in these solutions with respect to $m_1$ and $m_2$ is due to our choice to integrate over $\alpha_2$ rather than $\alpha_1$ when going from the Feynman parametrization \eqref{eq:bubble_feynman} to the one-fold integral \eqref{eq:bubble_2d_int}. 

\begin{figure}[t]
	\centering
	\includegraphics{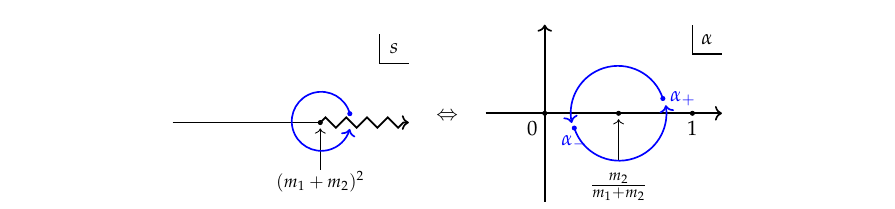}
    \caption{Analytic continuation of the $D=2$ massive bubble integral around the threshold singularity $s=(m_1+m_2)^2$, and the simultaneous path taken by the roots $\alpha_\pm$ as per equation \eqref{eq:bubble_2d_roots}. }
	\label{fig:bubble_2d_thresh}
\end{figure}
We now track what happens as we vary $s$ around the singular point $s = (m_1+m_2 )^2 $, as depicted in Figure \ref{fig:bubble_2d_thresh}. As $s$ performs a full rotation about the branch point, the roots $\a_+$ and $\a_-$ only rotate half-way around the point $m_2 / (m_1+m_2)$ due to the square root, and end up swapping positions.
In particular, because of the small $i \varepsilon$ part the roots $\alpha_\pm$ inherited from the Feynman $i \epsilon$ prescription, these roots end up crossing the original integration contour. This means the analytically-continued integration contour differs from the original one as follows:
\begin{equation}
    \begin{tikzpicture}[scale=1,baseline={([yshift=-.5ex]current bounding box.center)}]
\draw[->, thick] (0,2) -- (2, 2);
   \filldraw (0,2) circle(1pt);
   \filldraw (0.4, 1.8) circle(1pt) node[below] {$\alpha_-$};
   \filldraw (1.4, 2.2) circle(1pt) node[right] {$\a_+$};
   \node at (-0.5, 2) {\large $\gamma$:};
   \node[right] at (3, 2) {Original Contour};

   \draw[thick] (0,0) -- (0.2, 0);
   \draw[thick] (0.2, 0) arc (180:360:0.2);
\draw[thick] (0.6, 0) -- (1.2, 0);
\draw[thick] (1.6, 0) arc (0:180:0.2);
\draw[->, thick] (1.6, 0) -- (2, 0);
   
   \filldraw (0,0) circle(1pt);
   \filldraw (0.4, 0) circle(1pt) node[above] {$\alpha_+$};
   \filldraw (1.4, 0) circle(1pt) node[below] {$\a_-$};
   \node at (-0.5, 0) {\large $\gamma'$:};
 \node[right] at (3, 0) {Analytically Continued Contour};
\end{tikzpicture} 
\end{equation}
The difference between the original and analytically continued integral is thus given by the so-called difference contour, which is just a pair of residue contours:
\begin{equation} \label{eq:difference_contour}
\begin{tikzpicture}[scale=1,baseline={([yshift=-.5ex]current bounding box.center)}]
   \filldraw (0.4, 0) circle(1pt);
   \node[below] at (0.4, -0.2) {$\alpha_-$};
   \draw[->, thick] (0.6, 0) arc (0:350:0.2);
   \filldraw (1.4, 0) circle(1pt);
   \node[below] at (1.4, -0.2) {$\a_+$};
   \draw[->, thick] (1.6, 0) arc (360:10:0.2);
   \node[left] at (0, 0) {\large $\gamma' - \gamma$ = };
\end{tikzpicture} .
\end{equation}
Hence, the discontinuity of the massive bubble integral \eqref{eq:bubble_2d_int} with respect to the threshold singularity is nothing more than a specific linear combination of the residues of the integrand:
\begin{align} \label{eq:threshold_disc}
\Disc_{s - (m_1+m_2)^2}(I^{2D}_{\bubble}) = & \int_{\g'}\frac{d \a}{F} - \int_\g \frac{d\a}{F} = \int_{\begin{tikzpicture}[scale=0.5, baseline={([yshift=-.5ex]current bounding box.center)}]
    \filldraw (0,0) circle (1pt);
    \draw [->](0.21,0.125) arc (30:360:0.25);   
\end{tikzpicture} \ \begin{tikzpicture}[scale=0.5, baseline={([yshift=-.5ex]current bounding box.center)}]
    \filldraw (0,0) circle (1pt);
    \draw [<-](0.25,0) arc (0: 330:0.25);   
\end{tikzpicture}} \frac{d\a}{F}\,\\ \label{eq:threshold_disc_residues}
\sim & 2 \pi i \left( \res_{\a = \a_+}\left(\frac{1}{F} \right) - \res_{\a = \a_-}\left(\frac{1}{F}\right) \right)\\
 = & \frac{4 \pi i}{\sqrt{(s - ( m_1+m_2)^2 ) (s- (m_1-m_2)^2 )}}\,. \label{eq:disc_bubble_res}
\end{align}
Note that it was important to keep track of the orientations of the residue contours in~\eqref{eq:difference_contour} to determine the correct relative sign in~\eqref{eq:threshold_disc}.

\begin{figure}[t]
	\centering
	\includegraphics{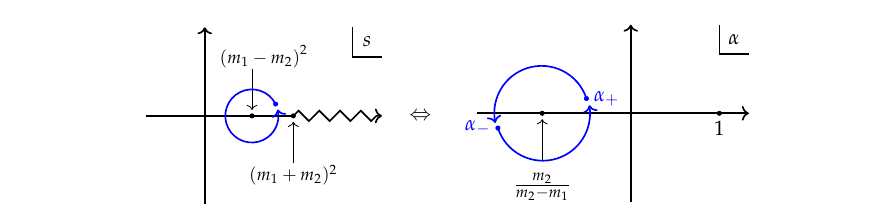}
    \caption{Analytic continuation of the $D=2$ massive bubble integral around the pseudo-threshold singularity $s=(m_1-m_2)^2$, and the simultaneous path taken by the roots $\alpha_\pm$ when  $m_1>m_2$, as per equation \eqref{eq:bubble_2d_roots}.}    \label{fig:bubble_2d_pseudo}
\end{figure}

We can also compute a discontinuity with respect to the pseudo-threshold singularity in~\eqref{eq:bubble_2d_pseudo}, by analytically continuing $s$ around $(m_1-m_2)^2$ as shown in Figure \ref{fig:bubble_2d_pseudo}. However, unlike for the threshold singularity, where the roots $\alpha_+$ and $\alpha_-$ intersected at a value $\frac{m_2}{m_1+m_2}$ that is between 0 and 1 for real masses, the roots now coincide at $\frac{m_2}{m_2-m_1}$. If $m_1 > m_2$, the latter intersection occurs on the negative real axis, while if $m_2 > m_1$ it occurs at a value greater than 1. In either case, the roots do not pinch the integration contour, which only runs between 0 and 1. As a result, the analytic continuation of $s$ leaves the integration contour invariant and the discontinuity is zero.

\subsection{$\alpha$-Positivity}
The fact that the threshold singularity gives rise to a branch cut on the physical Riemann sheet while the pseudo-threshold singularity does not can be understood as just the first instance of a general result:
\begin{shade}
\bigskip
Feynman integrals only have branch points that correspond to $\a$-positive singularities when considered in both the physical region and on the physical sheet.
\end{shade}
\noindent By \emph{physical region}, we mean values of the external kinematic values that describe physically-allowed configurations of particle scattering in Minkowski spacetime signature, and by \emph{physical sheet} we mean the Riemann sheet that we land on when we use the Feynman $i \varepsilon$ prescription. This result just amounts to the statement that before we have analytically continued our integral, the integration contour over each Feynman parameter extends from $0$ to $\infty$ along the positive real axis; thus, if the singularity occurs a finite distance away from the positive real axis in any of the $\alpha_e$ planes, the contour will not encounter it. Note that when applying this criteria, we need to check the complete set of Feynman parameters before the projective redundancy between these parameters is resolved (meaning, before we integrate over the delta function in~\eqref{eq: Feynman parametrization}). In the example just considered, this means we should require that both $\alpha_1$ and $\alpha_2 = 1 - \alpha_1$ are non-negative. An analogous result also holds for cosmological correlator integrals (which we recall take the form~\eqref{eq:cosmological_correlators}), whose initial integration contour in the $x_i$ variables also extends from $0$ to $\infty$ along the positive real axis.


\section{Discontinuities of Discontinuities}\label{sec:seqdisc}
We will now explore what happens when we iterate this procedure, to compute sequences of discontinuities of Feynman integrals. Determining which sequential discontinuities can occur for a given integral is especially important in perturbative bootstrap approaches, since it reveals a great deal about the class of functions to which the integral can evaluate. As we will see in this lecture, we can already derive strong constraints on what discontinuities can appear by considering how basic physical principles are encoded in Feynman integrals.

To this end, we first continue our study of the massive bubble integral, finding that discontinuities that are absent from the physical Riemann sheet can nevertheless arise after discontinuities have been computed with respect to other singularities. We then connect this phenomenon to the general framework provided by Picard–Lefschetz theory, and highlight some of the universal features of integrals over rational functions. In the second part of the lecture, we introduce the hierarchical principle and show how it can be used to show that certain sequences of discontinuities cannot occur. At the end, we will give a quick overview of additional types of analytic constraints that are obeyed by Feynman integrals.

\subsection{Bubble Integral Continued}
\begin{figure}[t]
	\centering
	\includegraphics{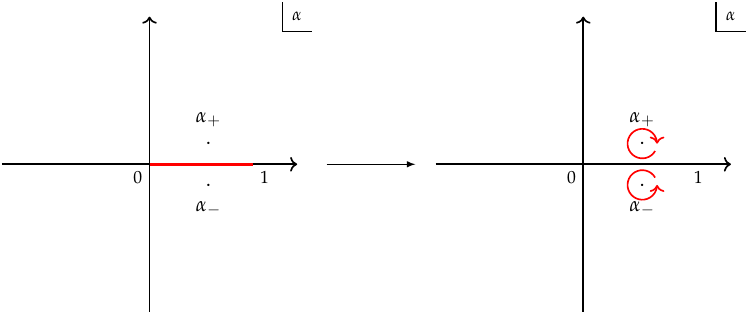}
    \caption{The computation of the discontinuity around the threshold $s=(m_1+m_2)^2$ amounts to changing the integration contour from the original one, shown on the left, to the residue contours shown on the right.}
	\label{fig:changeOfContour}
\end{figure}

Let us return to the example of the bubble integral in order to illustrate some of the features of iterated discontinuities. In the previous section we saw that the discontinuity of the bubble integral with respect to the pseudo-threshold $s=(m_1-m_2)^2$ is zero, in accordance with the $\alpha$-positivity criterion. We also saw that the discontinuity of the integral with respect to the threshold singularity $s=(m_1+m_2)^2$ can be expressed as an integral involving the same integrand but a different integration contour, namely
\begin{equation} \label{eq:disc_int}
    \Disc_{s - (m_1+m_2)^2}(I^{2D}_{\bubble}) = \int_{\gamma'-\gamma} \frac{d\alpha}{F} \, ,
\end{equation}
where the contour $\gamma'-\gamma$ was characterized in~\eqref{eq:difference_contour}. We have drawn the original contour and this new contour again in Figure \ref{fig:changeOfContour}. 

\begin{figure}[t]
	\centering
	\includegraphics{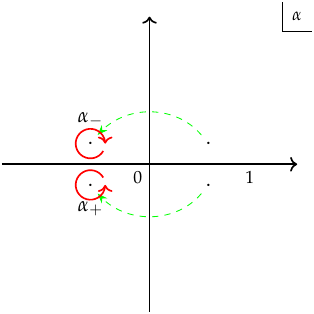}
    \caption{As the integral is analytically continued by sending $s$ towards $(m_1-m_2)^2$, the roots $\alpha_\pm$ drag the integration contour with them and meet on the negative real axis (for $m_1>m_2$).}
	\label{fig:Dragging of the new contour}
\end{figure}

\begin{figure}[t]
	\centering
	\includegraphics{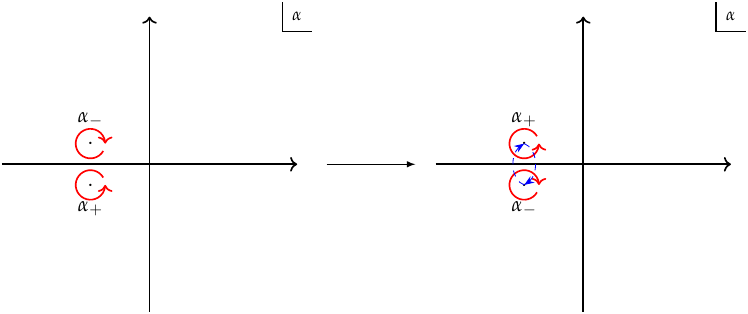}
    \caption{Upon continuing $s$ around $(m_1-m_2)^2$ the roots $\alpha_+$ and $\alpha_-$ swap position, so that the analytically-continued contour is the same as before except with the opposite orientation.}
	\label{fig:Factors of 2}
\end{figure}

With this in mind, we can think about computing a pseudo-threshold discontinuity of the integral in~\eqref{eq:disc_int}. Since this integral involves a different integration contour than the original Feynman integral, the fact that the roots $\alpha_+$ and $\alpha_-$ do not intersect on the real axis between 0 and 1 when $s \to (m_1-m_2)^2$ no longer implies the pseudo-threshold discontinuity will be zero. In fact, the new integration contour is localized to the roots $\alpha_+$ and $\alpha_-$ themselves; as we move these roots around, the integration contour will get dragged with them. Thus, wherever these roots collide they will necessarily pinch the contour, from which it follows that there is now a genuine singularity at the corresponding location in the $s$ plane. We have depicted how $\alpha_+$ and $\alpha_-$ move and drag the contours as $s \to (m_1-m_2)^2$ in Figure \ref{fig:Dragging of the new contour} (for the case where $m_1>m_2$).  
To compute the pseudo-threshold discontinuity of~\eqref{eq:disc_int}, we carry out the same steps as when we computed the discontinuities of the original Feynman integral. Continuing $s$ around $(m_1-m_2)^2$, we find that the roots $\alpha_+$ and $\alpha_-$ swap position. Correspondingly, the analytically-continued contour can be seen to be nothing other than the same residue contours, but now with the opposite orientation. This is depicted in Figure \ref{fig:Factors of 2}. It follows that the difference contour is twice the analytically-continued contour:
\begin{align} \label{eq:double_disc}
\begin{split}
    &\text{Disc}_{s-(m_1-m_2)^2}(\text{Disc}_{s-(m_1+m_2)^2}(I^{2D}_{\begin{tikzpicture}[scale=0.1]
    \draw (-2, 0) -- (-1, 0);
    \draw (1, 0) -- (2, 0);
    \draw (0,0) circle (1);
\end{tikzpicture}}))=\\
    &=-\text{Disc}_{s-(m_1+m_2)^2}(I^{2D}_{\begin{tikzpicture}[scale=0.1]
    \draw (-2, 0) -- (-1, 0);
    \draw (1, 0) -- (2, 0);
    \draw (0,0) circle (1);
\end{tikzpicture}})-\text{Disc}_{s-(m_1+m_2)^2}(I^{2D}_{\begin{tikzpicture}[scale=0.1]
    \draw (-2, 0) -- (-1, 0);
    \draw (1, 0) -- (2, 0);
    \draw (0,0) circle (1);
\end{tikzpicture}})=\\
    &=-2[\text{Disc}_{s-(m_1+m_2)^2}(I^{2D}_{\begin{tikzpicture}[scale=0.1]
    \draw (-2, 0) -- (-1, 0);
    \draw (1, 0) -- (2, 0);
    \draw (0,0) circle (1);
\end{tikzpicture}})]=\\
    &=\frac{-8\pi i}{\sqrt{(s-(m_1+m_2)^2)(s-(m_1-m_2)^2)}}.
    \end{split}
\end{align}

\subsection{Picard--Lefschetz theory}
While everything can be explicitly worked out in the bubble example, tracking the requisite analytic continuations becomes increasingly complicated when the number of integration variables increases. Thus, it is useful to generalize our discussion to a more abstract framework for computing these discontinuities. This framework is provided by \textit{Picard-Lefschetz theory}. 

In the computation of both the threshold and pseudo-threshold discontinuities, we have observed that the final results can be cast as integrals over the same rational function, but with a new integration contour. This is in fact a general phenomenon, and holds whenever one computes any sequence of discontinuities of an integral over a rational function. This is formalized by the Picard-Lefschetz formula, which states that
\begin{shade}
\begin{align} \label{eq:picard_lefschetz}
    \text{Disc}_{\lambda}\left(\int_{\gamma}\dd I\right)&=c\int_{\gamma'}\dd I\,,
\end{align}
\end{shade}
\noindent where $\gamma'$ is called the \textit{vanishing cycle}, because it vanishes when $\lambda=0$, and $c$ is an (oriented) intersection number between the original contour $\gamma$ and the \textit{vanishing cell}, which is just a dual contour that ends on the two singularity surfaces that coincide when $\lambda=0$.

\begin{figure}[t]
	\centering
	\includegraphics{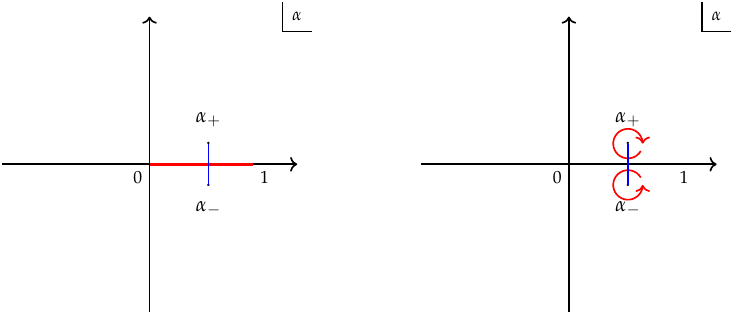}
    \caption{On the left, the vanishing cell associated with the threshold singularity (depicted in blue) intersects the original Feynman integration contour (depicted in red) precisely once. On the right, the same vanishing cell (again in blue) intersects the new integration contour (again in red) twice.}
	\label{fig:picard-lefshetz}
\end{figure}

Note that the existence of a contour $\gamma'$ that computes the discontinuity for us follows simply from the fact that we are computing the difference between two integrals that can be analytically continued into each other. Since our integrands are rationals, they don't change as long as we analytically continue along closed loops; it is only the contours that change.  
However, the Picard-Lefschetz formula tells us a bit more; it informs us that the difference contour has the property that it becomes trivial, or vanishes, as $\lambda\rightarrow0$. Note that this is consistent with our threshold and pseudo-threshold examples. In both cases, the residue contours that we found were pairs of loops that were oriented in opposite directions, which cancel against each other if we think of the roots $\alpha_+$ and $\alpha_-$ as sitting on top of each other. Moreover, in both of these examples, the vanishing cell was nothing more  than the line from $\alpha_+$ to $\alpha_-$ that can be shrunk to length zero in the corresponding singular limit. When we computed the threshold singularity of our original Feynman integral, this line intersected the integration contour precisely once, so $c=1$; conversely, when we computed the pseudo-threshold discontinuity of our original Feynman integral, this line did not intersect the integration contour, so $c=0$. However, when we computed the pseudo-threshold discontinuity of the threshold discontinuity, this line now intersected the contour twice, meaning $c=2$. This is precisely the factor of 2 that we got from the doubling of the contour in~\eqref{eq:double_disc}. We have illustrated what the vanishing cell looks like in two examples in Figure \ref{fig:picard-lefshetz}.

\subsection{The Hierarchical Principle}\label{ssec:hierprinciple}

Unfortunately, there doesn't exist an efficient method for determining the explicit form of the vanishing cycle $\gamma'$ or the value of $c$ in complicated examples. However, as we will see shortly, the Picard-Lefschetz formula still tells us enough to allow us to make nontrivial predictions about Feynman integrals. 

To motivate the types of constraints we will introduce, let's first recall Cutkosky's formula \cite{Cutkosky:1960sp}, which gives us an integral formula for computing the first discontinuities of Feynman integrals. In particular, if we want to compute the discontinuity with respect to a singularity at $\lambda = 0$ that is associated with a unique Landau diagram $G'$ (which we recall is simply the graph that one gets by contracting all the propagators in $G$ that are off-shell in the corresponding solution to the Landau equations), this formula takes the form
\begin{align} \label{eq:cutkosky}
\Disc_{\lambda}(I_G) &= \int \prod_{i=1}^L \dd^D l_i \frac{\prod_{e \in G'} \de^+ (q_e^2-m_e^2)}{\prod_{e \notin G'} (q_e^2-m_e^2)}\,, 
\end{align}
where the $\delta^+$ functions are defined by
\begin{equation}
\de^+ (q_e^2-m_e^2) = \de (q_e^2-m_e^2)\Theta (\pm q_e^0)\,,
\end{equation}
and the sign that appears in each theta function is chosen so that the energy always flows in a consistent direction through the graph $G'$. More generally, when there exist multiple solutions to the Landau equations that give rise to a singularity at the same location in the space of external kinematics (and thus there exist multiple Landau diagrams $G'$ we can associate with this singular location), we will need to sum over multiple integrals of the form~\eqref{eq:cutkosky}. 

In words, Cutkosky's formula tells us that the first discontinuities of Feynman integrals can be computed by integrating over all real values of the loop momenta, after we have placed on shell all of the propagators that are involved in pinching the integration contour. Often this is referred to as \emph{cutting} these propagators; it just amounts to trading our original integration contour for one that computes a number of residues of the integrand, similar to what we have been finding in Feynman parameter space.  The formal proof of Cutkosky's formula was given by Pham \cite{pham1967introduction}:
\begin{shade}
\bigskip
Theorem (Pham): the discontinuity of a Feynman integral $I_G$ with respect to $\a$-positive singularities in which each loop in $G$ has at least one nonzero $\a$ is given by Cutkosky's formula. 
\end{shade}
\noindent Note that the restriction to singularities that occur when at least one Feynman parameter in each loop takes a nonzero value is actually quite weak, as a basis of integration contours can always be chosen such that at least one propagator is on shell in every loop. Unfortunately, no closed formula analogous to Cutkosky's formula yet exists for longer sequences of discontinuities. 

The key point we want to highlight in Cutkosky's formula is that the integral~\eqref{eq:cutkosky} vanishes off of the support of the $\delta$ functions that place all of the propagators in $G'$ on shell. Therefore, any solution to the Landau equations (for the original diagram $G$) that requires one or more of these propagators to be off-shell can no longer be kinematically reached in the new integral. We can formalize this by defining a modified set of Landau equations, that hold after we have computed a discontinuity with respect to the singularity at $\lambda =0$ (which, for simplicity, we again assume arises in connection with a unique Landau diagram $G'$):
\begin{shade}
\begin{align}
    \label{eq:landua_mod_1}q_e^2-m_e^2+i\varepsilon &= 0 \qquad  \text{for every edge $e \in G'$}\\
    \label{eq:landua_mod_2}\a_e(q_e^2-m_e^2+i\varepsilon) &= 0 \qquad  \text{for every edge $e \not\in G'$}\\
    \label{eq:landua_mod_3}\sum_{e\in{\rm loop}} \pm\a_eq_e^\mu &= 0 \qquad  \text{for every independent loop in $G$.}   
\end{align}
\end{shade}
\noindent If there exists a solution to the original Landau equations at $\lambda' = 0$, but there no longer exists any such solution to these modified equations, we can conclude that
\begin{equation}
\mathrm{Disc}_{\lambda'}\mathrm{Disc}_{\lambda}\left(I_G\right)=0\,,
\end{equation}
since the second discontinuity must be zero if there is no longer any singularity at $\lambda'=0$.

The \textit{hierarchical principle}, first articulated in the 1960s~\cite{Landshoff1966,pham1967introduction,boyling1968homological}, extends this result to all sequential discontinuities of Feynman integrals. Loosely speaking, this principle states that once we have computed a discontinuity in which a given propagator has played a role in pinching the integration contour, this propagator is now permanently set on shell, no matter how many subsequent discontinuities we compute. Intuitively, this principle articulates nothing more than what we have already seen in Feynman parameter space in the example of the bubble integral: once the integration contour has localized to a given pole in the integrand, no amount of analytic continuation will produce a contour that un-moors itself from this singular surface.

\begin{figure}[t]
	\centering
	\includegraphics{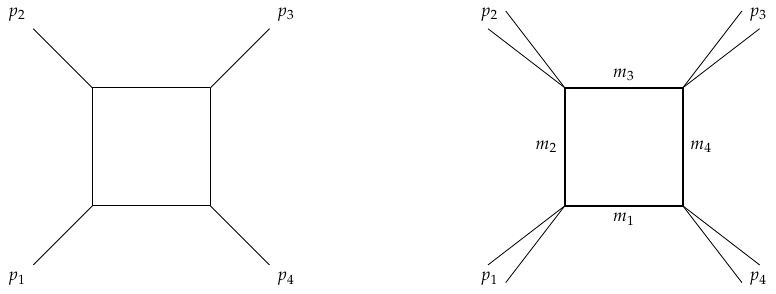}
    \caption{\emph{Left:} the massless box integral, which involves both massless propagators and massless external particles. \emph{Right:} the totally massive box integral, in which all external and internal particles are massive, with distinct masses.}
	\label{fig:box3}
\end{figure}

\begin{figure}[t]
	\centering
	\includegraphics{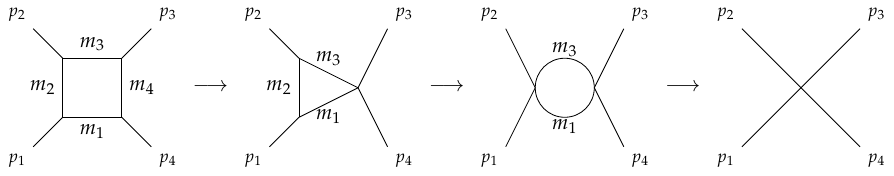}
    \caption{An example of a hierarchically-allowed sequence of Landau diagrams associated for the massive box integral. If we read this sequence from right to left, we can interpret each arrow as representing a discontinuity. If these diagrams appeared in any other order, the corresponding sequence of discontinuities would vanish.}
	\label{fig:hierarchical_principle_ex}
\end{figure}

Although this principle goes back sixty years, very little use has been made of it in practice. This is because its application is extremely subtle. We can see this by comparing what it implies for the box integral with and without masses. In particular, let us consider the totally generic box integral (that has nonzero internal and external masses, all of which are distinct), as well as the totally massless box; these two examples are depicted in Figure \ref{fig:box3}.
In the case of the totally massive box, there is just one singularity associated with each possible set of on-shell propagators; the singularities of this integral are thus in one-to-one correspondence with each of the Landau diagrams that one can construct. This makes it easy to enumerate the sequences of discontinuities that are consistent with the hierarchical principle; they will just be those that correspond to sequences of Landau diagrams in which no propagator is ever taken off-shell. An example of an allowed sequence of Landau diagrams is shown in Figure~\ref{fig:hierarchical_principle_ex}. In contrast, if we consider the pair of threshold singularities that one gets by putting either the horizontal or the vertical propagators in Figure~\ref{fig:box3} on shell, we have that
\begin{equation}
\mathrm{Disc}_{s-(m_1+m_3)^2}\mathrm{Disc}_{t-(m_2+m_4)^2}\left(I_G\right) = 0\,,
\end{equation}
since no singularity arises at $s=(m_1+m_3)^2$ after the propagators that have mass $m_2$ and $m_4$ are placed on shell.

\begin{figure}[t]
	\centering
	\includegraphics{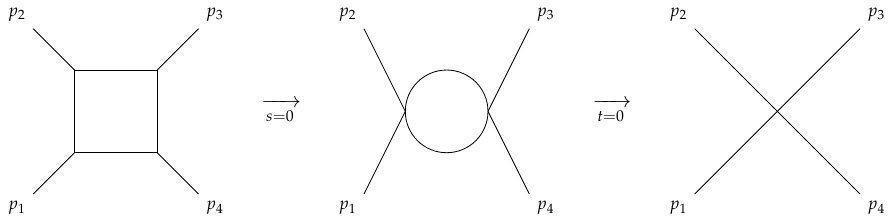}
    \caption{Example of a sequence of discontinuities that can appear in the massless box integral. Even though, na\"ively, the singularities at $s=0$ and $t=0$ are associated with disjoint sets of on-shell propagators, $\mathrm{Disc}_{s}\mathrm{Disc}_{t}\left(I_G\right)$ can be nonzero when $s=0$ and $t=0$ singularities also exist on the support of all four on-shell constraints.}
	\label{fig:hierarchical_principle_massless_ex}
\end{figure}

Once we start setting the masses in the box integral to be equal or zero, singularities start colliding. For instance, the singularity that corresponds to putting all four propagators on shell in the generic box integral occurs at
\begin{equation}         
m_1^4m_3^4+m_2^4m_4^4-2m_1^2m_2^2m_3^2m_4^2-2st(m_1^2m_3^2+m_2^2m_4^2)+s^2t^2=0\,,
\end{equation}
where $s=(p_1+p_2)^2$ and $t=(p_2+p_3)^2$.  
However, if we set all internal masses to be zero, namely $m_1=m_2=m_3=m_4=0$, this becomes  
\begin{equation}
    s^2 t^2=0\,.
\end{equation}
Thus, in the massless case we expect singularities at $s=0$ and $t=0$ to be accessible even when all four propagators are on shell. This means that, for the massless box, we can have
\begin{equation} \label{eq:box_double_disc}
\mathrm{Disc}_{s}\mathrm{Disc}_{t}\left(I_G\right) \neq 0\,.
\end{equation}
because this double discontinuity can be associated with the sequence of Landau diagrams shown in Figure~\ref{fig:hierarchical_principle_massless_ex}.\footnote{For more examples like this, and for more background on the types of diagrams shown in Figures~\ref{fig:hierarchical_principle_ex} and \ref{fig:hierarchical_principle_massless_ex}, see~\cite{pham1967introduction,Hannesdottir:2022xki,Hannesdottir:2025bss}.}

Thus, making use of the hierarchical principle in practice generally requires having knowledge of \emph{all} solutions to the Landau equations. At this point in time, no technology exists for efficiently finding all solutions to these equations, due to the need for algebraic blowups~\cite{Berghoff:2022mqu,Fevola:2023fzn,Vergu:2025mag}. However, we will now turn to a method that will allow us to sidestep this problem, and make nontrivial predictions even when we are not sure all solutions to the Landau equations have been found.

\subsection{Hierarchical Predictions in Practice}
Let us imagine we already have a list of external kinematic loci where we believe a Feynman integral may become singular, even though we don't know where in the space of integration variables these singularities arise. Following~\cite{pham1967introduction,Fevola:2023fzn}, it turns out that we can confirm these loci are singular in the span of integrals that are associated with this Feynman graph, without explicitly invoking the Landau equations.

\begin{figure}[t]
	\centering
	\includegraphics{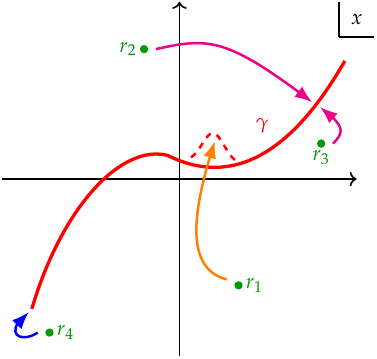}
    \caption{Illustration of the different ways singularities in the integrand can intersect the integration contour. In orange, we depict a singularity that intersects the original integration contour, but that the contour can be deformed away from. In pink, we see a pinch singularity, and in blue we see an endpoint singularity.}
	\label{fig:sing3_again}
\end{figure}

Consider again the figure we drew in the first lecture (redrawn in Figure \ref{fig:sing3_again}), illustrating how singularities in the integrand can be inherited by the full integral. In general, this only happens when the singular surfaces in the integrand intersect an integration endpoint, or pinch the integration contour. Let us think about the consequences this holds for Feynman integrals in the \textit{Lee-Pomeransky representation}~\cite{Lee:2013hzt}, which takes the form
\begin{equation}\label{eq:lee_pom}
I=\int_{\alpha_e\geq0}\frac{\dd\alpha_1\ldots\dd\alpha_E}{\mathcal{G}^{D/2}}\,,
\end{equation}
where $\mathcal{G}$ is a polynomial in the Feynman parameters $\alpha_e$ and Lorentz-invariant kinematic variables. (We will not derive this representation here; for more information, see for instance~\cite{Weinzierl:2022eaz}.)
We can naturally associate a topological space to any integral that takes this form, by taking the space $\mathbb{C}^E$ that the $\alpha_e$ variables live in, and deleting the points that satisfy $\mathcal{G}=0$ (which is where the integrand is not well-defined) as well as the integration endpoints at $\alpha_e=0$:
\begin{equation}
    X=\mathbb{C}^E\setminus\left\{G=0 \vee \,\prod_{e} \alpha_e=0\right\}\,.
\end{equation}
The key observation is that the topology of the space $X$ will change if we go to either a pinch or endpoint configuration. In a pinch configuration the hypersurface $\mathcal{G}=0$ becomes (more) degenerate, while in an endpoint configuration the $\mathcal{G}=0$ surface intersects one of the $\alpha_e=0$ hyperplanes. 

In practice, we can probe for such changes of topology using the (signed) Euler characteristic $|\chi(X)|$, which constitutes a topological invariant that can be efficiently computed (at a given kinematic point) using modern algebraic geometry methods. We do this by first computing the Euler characteristic at a generic kinematic point, and then computing it at a kinematic point that satisfies $\lambda = 0$, where $\lambda=0$ is one of our candidate singular loci. We then ask:
\begin{equation} \label{eq:euler_test}
    \abs{\chi(X|_{\lambda=0})} \stackrel{?}{<} \abs{\chi(X)} \,.
\end{equation}
If these two numbers are the same, we interpret that to mean that the topology has not changed, which means we cannot possibly be at a pinch or endpoint configuration.\footnote{In principle, one could imagine encountering a degenerate situation in which the topology changes but the signed Euler characteristic remains the same; in practice, this does not seem to happen for Feynman integrals.} Conversely, if the signed Euler characteristic drops in the $\lambda \to 0$ limit, this means that one or more of the bounded cells in $X$ has vanished. Thinking about the set of inequivalent integration contours that can be defined on $X$, these vanishing cells signal the existence of vanishing cycles that can enter the Picard-Lefschetz formula~\eqref{eq:picard_lefschetz}.

Now let's go a step further, and combine the Euler characteristic test~\eqref{eq:euler_test} with the hierarchical principle.
Think back to what happened when we computed the threshold discontinuity of the bubble integral. As we analytically continued $s$ around $(m_1+m_2)^2$ we saw that the singularities of the integrand got tangled up with the integration contour, such that the difference contour $\gamma'-\gamma$ in~\eqref{eq:difference_contour} completely localized to these singularities of the integrand.  
The important detail to notice for the present discussion is that the new integration contour no longer ends at either $\alpha=0$ or $\alpha=1$. This is in fact a general phenomenon, which generalizes to the Lee-Pomeransky representation~\eqref{eq:lee_pom}: whenever a singularity arises from a pinch of the $\alpha_e$ integration contour, the new (difference) contour will not end at the original integration endpoint $\alpha_e=0$~\cite{Berghoff:2022mqu}. Intuitively, this is simply because the integration cycle that encodes the discontinuity in the Picard-Lefschetz formula must be contained within an infinitesimal ball around the singular surface---so if this pinch happens away from the origin, the $\alpha_e=0$ endpoint can no longer be involved. 

This ``forgetting of integration endpoints'' is the parametric version of the hierarchical principle. Just as propagators permanently go on shell in the momentum space representation, endpoints of the integration contour are permanently forgotten in parametric representations. Recall that this information about the endpoints of our current integration contour also explicitly entered the definition of our space $X$. Thus, the obvious move is to define a modified set of spaces
\begin{equation}
    X_P=\mathbb{C}^E\setminus\left\{G=0 \vee \,\prod_{e\notin P}\alpha_e=0\right\}\,,
\end{equation}
where $P$ keeps track of the set of $\alpha_e$ integration contours that no longer end at the origin, since they have been pinched by previous discontinuities. This modified space then allows us to ask whether there still exists a singularity at a given kinematic location $\lambda = 0$ after these discontinuities have been computed. Namely, we ask
\begin{equation}
    \abs{\chi(X_P|_{\lambda=0})} \stackrel{?}{<} \abs{\chi(X_P)}, ,
\end{equation}
and interpret the result as before.

\begin{figure}[p]
	\centering
	\includegraphics{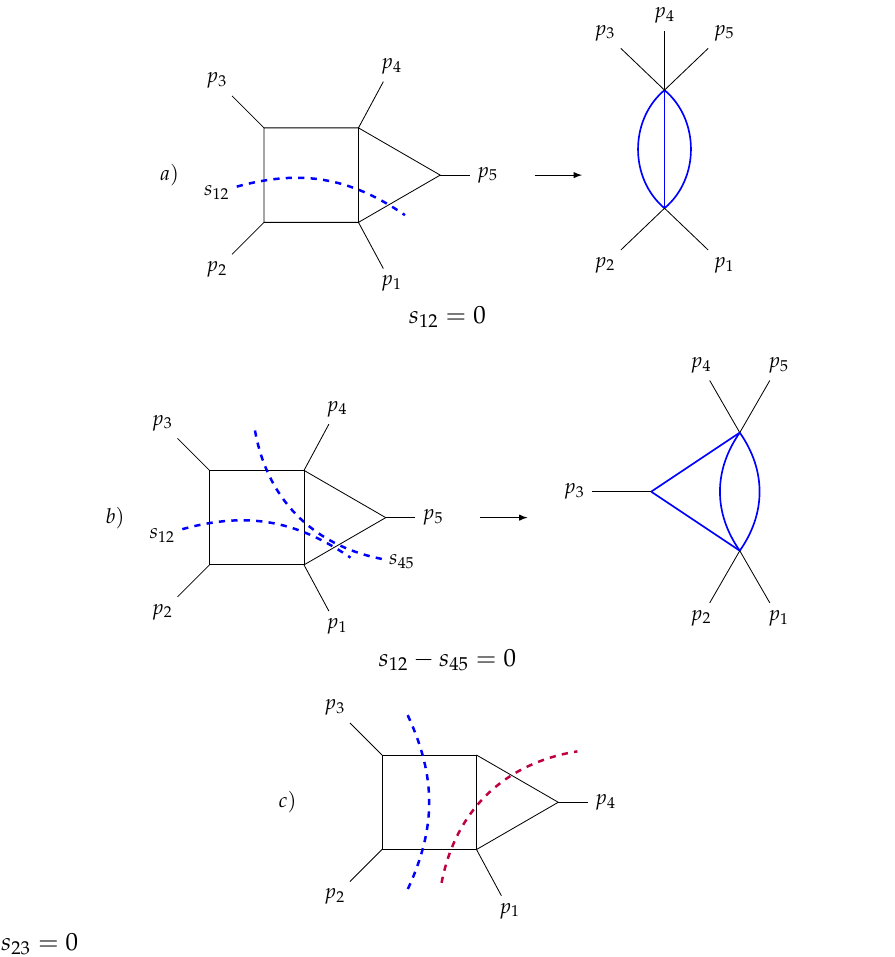}
    \caption{$a$) The minimal cut for $\lambda=s_{12}$ in the diagram on the left must partition the diagram shown into two parts, such that the sum of momenta $p_1+p_2$ is resolved. This requirement gives rise to the minimal cut shown on the right. $b$) If we instead consider the singularity at $\lambda = s_{12}-s_{45}$, our minimal cut must resolve both $p_1+p_2$ and $p_4+p_5$ separately, as shown. $c$) If external momenta do not enter every vertex of our Feynman diagram, there may be multiple minimal cuts, as shown in this example in red and blue.}
	\label{fig:minimal_cut}
\end{figure}

In order to use this idea in practice, there remains just one more question that must be addressed: how do we know which Feynman parameter contours have been pinched by discontinuities that have already been computed? For instance, if all one knows is that a discontinuity with respect to $\lambda'=0$ has been computed, then in general one has no information about which Feynman parameter contours were pinched versus in endpoint configurations. (In fact, the answer to this question will depend on which Riemann sheet this discontinuity was computed on.) However, it turns out there is an easy graphical way to provide a conservative answer to this question, using \textit{minimal cuts}~\cite{Hannesdottir:2024cnn}.  
Given a Feynman diagram and a singular kinematic surface $\lambda(\{s_{i,\dots,j}\},\{m^2_k\})=0$ we refer to a set of cut propagators as minimal if 
\begin{enumerate}[label=(\roman*)]
    \item the cut propagators partition the external momenta into the combinations that appear in the Mandelstam variables $\{s_{i,\dots,j}\}$, where $s_{i,\dots,j} = (p_i+\dots+p_j)^2$;
    \item each of the internal masses $\{m^2_k\}$ appears in at least one of the cut propagators (unless this mass has already appeared as one of the Mandelstam variables);
    \item one of the first two properties is no longer satisfied if any of the cut propagators are taken off shell.
\end{enumerate}
Intuitively, minimal cuts just identify the minimal set(s) of propagators that must be placed on shell in order for the corresponding Landau diagram to resolve the relevant kinematic surface. For instance, if we are interested in the surface $\lambda=(p_1+p_2)^2 = 0$, then we need to put a sufficient set of propagators on shell in order to ensure that a third external momentum $p_3$ doesn't also enter the corresponding Landau diagram at the same vertex as $p_1$ and $p_2$; if we do not, this diagram will only be sensitive to the linear combination of momenta $p_1+p_2+p_3$. Several examples of minimal cuts are shown in Figure \ref{fig:minimal_cut}.

\FloatBarrier

With all these tools (finally) at our disposal, we can now place hierarchical constraints of the form
\begin{equation}\label{eq: generic constraint}
\mathrm{Disc}_{\lambda_2}\ldots\mathrm{Disc}_{\lambda_1}\ldots\left(I_G\right)=0
\end{equation}
on the Feynman integrals that are associated with the graph $G$. First, we construct the minimal cut(s) that correspond to $\lambda_1=0$. This tells us what space $X_P$ we should use; namely, we take $P$ to be the set of propagators in the minimal cut, whose $\alpha_e$ contours will have been pinched when $\mathrm{Disc}_{\lambda_1}$ was computed (if there are multiple minimal cuts, we should consider the space $X_P$ associated with each of them). We then check whether $\abs{\chi(X_P|_{\lambda_2=0})} < \abs{\chi(X_P)}$. If these two numbers are the same, we conclude that there can be no pinch or endpoint singularity at $\lambda_2=0$ after we have computed $\mathrm{Disc}_{\lambda_1}$, so \eqref{eq: generic constraint} describes a valid constraint. We emphasize that this test explicitly incorporates knowledge about the topology of the Feynman graph $G$, so these constraints only apply to integrals with the same topology (or its contractions). 


\subsection{Example: the Two-Mass Easy Box}

\begin{figure}[t]
	\centering
	\includegraphics{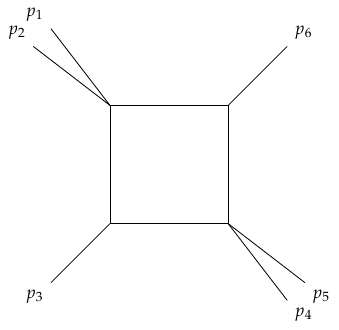}
    \caption{The two-mass easy box.}
	\label{fig:easybox}
\end{figure}

\begin{figure}[t]
	\centering
	\includegraphics{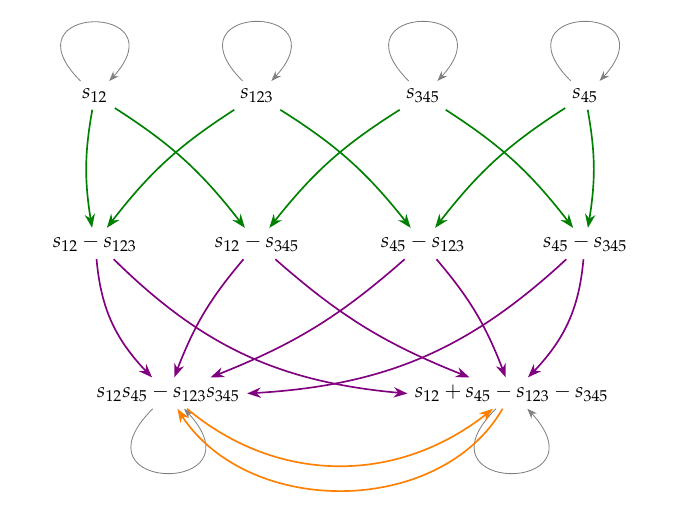}
    \caption{Sequences of discontinuities that are allowed by the hierarchical principle, in the example of the two-mass easy box.}
	\label{fig:hierarchy}
\end{figure}

As an example, let us consider the scalar ``two-mass easy box'' integral, shown in Figure \ref{fig:easybox}. This integral has 10 singular kinematical loci, which occur where the polynomials $\{s_{12}, s_{123}, s_{45}, s_{345}, s_{12}-s_{123}, s_{12}-s_{345}, s_{45}-s_{123}, s_{45}-s_{345}, s_{12} s_{45}-s_{123} s_{345}, s_{12}+s_{45}-s_{123}-s_{345}\}$ vanish. Carrying out the above analysis for all pairs of singularities, one finds that no fewer than 64 constraints of the form~\eqref{eq: generic constraint} can be derived~\cite{Hannesdottir:2024cnn}. The sequences of discontinuities that are \emph{allowed} by these constraints are indicated by the arrows in Figure \ref{fig:hierarchy}. For instance, this tells us that 
\begin{equation}
    \mathrm{Disc}_{s_{12}-s_{345}} \mathrm{Disc}_{s_{12}-s_{123}}\mathrm{Disc}_{s_{12}} (I_G) = 0
\end{equation} 
for the two-mass easy box, since there is no sequence of arrows that allows us to move from $s_{12}-s_{123}$ to $s_{12}-s_{s345}$. It turns out that one can also show, using similar ideas, that repeated discontinuities with respect to any of the singularities $\{s_{12}-s_{123}, s_{12}-s_{345}, s_{45}-s_{123}, s_{45}-s_{345}\}$ (those that appear in the middle row of Figure \ref{fig:hierarchy}) also disappear. In other words, $\mathrm{Disc}_{s_{12}-s_{s345}}\cdots \mathrm{Disc}_{s_{12}-s_{s345}}\cdots (I_G) = 0$, and similarly for the other indicated singularities. This is reflected in Figure~\ref{fig:hierarchy} by the absence of circular arrows that start and end on these singularities (while there are circular arrows for all of the other singularities, with respect to which repeated discontinuities can be computed). Deriving these further constraints requires developing a little more machinery, so we will not do so here. However, these constraints, along with all of the other constraints encoded in Figure~\ref{fig:hierarchy}, can be checked to all orders in dimensional regularization.  

\subsection{Other Types of Constraints}
In addition to the $\alpha$-positivity and hierarchical principles, there exist constraints on which discontinuities can appear immediately after one another in scattering amplitudes. The most prominent class of such constraints are the \textit{Steinmann relations} \cite{steinmann1960zusammenhang,Cahill:1973px,Caron-Huot:2016owq,Benincasa:2020aoj}, which tell us that double discontinuities in partially-overlapping momentum channels must vanish:  
\begin{equation}\label{eq:steinmann}
    \mathrm{Disc}_{s_J-M_J}\left(\mathrm{Disc}_{s_I-M_I}\left(\mathcal{A}\right)\right)=0\,,
\end{equation}
where $I$ and $J$ represent partially-overlapping sets of external momentum indices such that
\begin{equation}\label{eq:partially_overlapping}
    \begin{aligned} 
        I\cap J &\neq\emptyset\,, \\
        I \not\subset J \quad \mathrm{and}&\quad J \not\subset I\,.
    \end{aligned}
\end{equation}
The corresponding Mandelstam invariants are given by
\begin{equation}
    S_I=\left(\sum_{i\in I}p_i\right)^2\,,
\end{equation}
while we have adopted the notation that
\begin{equation}
    M_I=\left(\sum_{e\in \mathrm{cut}}m_e\right)^2\,
\end{equation}
denotes the squared sum of internal masses that appear in the propagators that are cut in the channel $I$. Note that, unlike for hierarchical constraints, where we are allowed to compute any number of discontinuities before and between the two that are incompatible, the Steinmann relations~\eqref{eq:steinmann} only constrain the first two discontinuities. 

\begin{figure}[t]
	\centering
	\includegraphics{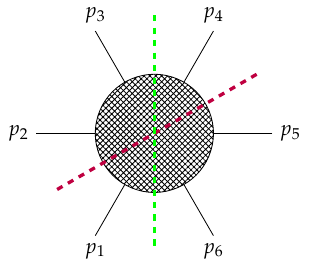}
    \caption{Incompatible sequence of discontinuities for a six-point process predicted by the Steinmann relations. The red cut represents the discontinuity along the $s_{234}=0$ singularity, while the green cut represents the discontinuity along the $s_{123}=0$ singularity.}
	\label{fig:six_particles_Steinmann_ex}
\end{figure}
\FloatBarrier

Although the Steinmann relations were originally only proven back in the sixties for theories with a mass gap \cite{steinmann1960zusammenhang}, they have also been observed to hold in massless theories, even in individual Feynman integrals. As an example, massless six-particle amplitudes $\mathcal{A}_6$ (and indeed, all the Feynman integrals that contribute to such amplitudes) have been observed to obey the constraint
\begin{equation}
\mathrm{Disc}_{s_{234}}\mathrm{Disc}_{s_{123}} \mathcal{A}_6 = 0\,,
\end{equation}
corresponding to the pair of overlapping cuts shown in Figure \ref{fig:six_particles_Steinmann_ex}. More generally, the only observed exceptions to this rule in massless integrals and amplitudes are instances involving massless two-particle channels. Indeed, in~\eqref{eq:box_double_disc} we saw that the totally massless box has a double discontinuity with respect to the partially-overlapping channels $s=s_{12}$ and $t=s_{23}$. However, in general, as long as both discontinuity channels involve at least three massless external momenta, the Steinmann relations have been observed to hold (for more on this topic, see~\cite{Hannesdottir:2025bss}).

In fact, further extensions of the Steinmann relations have been observed in the context of planar $\mathcal{N}=4$ super-Yang-Mills theory. Namely, in the integrals that contribute to this theory, the Steinmann relations also hold after any number of additional discontinuities have been computed:
\begin{equation}\label{eq:extended_steinmann}
\mathrm{Disc}_{s_J}\mathrm{Disc}_{s_I} \cdots \left(I_G \right)=0\,,
\end{equation}
where, in addition to requiring the sets $I$ and $J$ to be partially-overlapping in the sense of~\eqref{eq:partially_overlapping}, we also require that $|I|>2$ and $|J|>2$ in order to avoid the types of exceptions mentioned in the last paragraph. These constraints go by the name of the \emph{extended Steinmann relations}~\cite{Caron-Huot:2019bsq}, and have interesting ties to cluster algebras~\cite{Drummond:2017ssj}.  

Steinmann-type constraints have also been found to apply in cosmology~\cite{Benincasa:2020aoj}, and (observationally) there exist many discontinuities that never appear in immediate succession in most Feynman integrals (where these restrictions only hold in a given integral, not across all integrals). It is hoped that robust methods will be developed in the coming years that allow us to predict these types of constraints, similar to how minimal cuts and the Euler characteristic test allow us to easily work out the implications of the hierarchical principle.


\section{Special Functions}\label{sec:special}
We now take a slight detour to consider the types of special functions that arise when we evaluate Feynman integrals, in or near integer numbers of dimensions (which is to say, order-by-order in the dimensional regularization parameter $\epsilon = (4-D)/2$). This will allow us to see more concretely how knowledge of the analytic structure of Feynman integrals allows us to place strong constraints on the functions that they evaluate to. For a more complete introduction to the ideas in this lecture, see~\cite{Duhr:2014woa,Bourjaily:2022bwx}.

\subsection{Integrals of Rational Functions}

Let us return again to the generic integral we began by studying in lecture~\ref{sec:sing}, namely the integral
\begin{equation}
f(z) = \int_\gamma  \frac{g(x, z)}{h(x,z)}d x\, ,
\end{equation}
where $g(x, z)$ and $h(x, z)$ are both polynomials. Given any such integral, we can always partial fraction the integrand with respect to $x$, to separate out terms that exhibit different dependence on the integration variable: 
\begin{equation}
\int \frac{g(x, z)}{h(x,z)} dx \to \int \Big[ \sum_i P_i(z) x^i + \sum_{i, a_j} \frac{R_{i,a_j}(z)}{(x - a_j(z))^i}\Big] dx 
\end{equation}
for some $P_i(z)$ and $R_{i,a_j}(z)$, where the $a_j(z)$ record the positions of all the singularities that appear in the integrand. Almost all of these terms can be integrated in terms of rational (or possibly algebraic) functions of $z$. The exception will be the terms that involve simple poles, which generate logarithms: 
\begin{equation}
\int \frac{\dd x}{x - a} = \log(x - a)\,.
\end{equation}
Thus, the first special functions that we expect to appear in Feynman integrals (which indeed start their lives as integrals over rational functions) are logarithms. However, we can ask: what type of functions might we expect to show up more generally? 

In the simplest cases, we only encounter simple poles even when we carry out many integrations. In such cases, Feynman integrals evaluate to so-called multiple polylogarithms (MPLs), which are defined recursively by
\begin{align} \label{eq:mpl_def}
G_{a_1, a_2, \cdots, a_n}(z) &\coloneqq \int_0^z \frac{G_{a_2, \cdots, a_n}(x)}{x - a_1} d x, \\
G_{\underbrace{0,\cdots, 0}_{n}}(z) &\coloneqq \frac{\log^n(z)}{n!}\, ,
\end{align}
where the second equation regulates the case where the general definition is divergent. The space of MPLs is graded by the number of integrations $n$ that appear in each function; this number is usually referred to as the \emph{transcendental weight} (or just \emph{weight}) of an MPL. The classical polylogarithms
\begin{equation} \label{eq:classical_polylog_def}
\Li_n(z) = \int_0^z \frac{\Li_{n-1}(x)}{x} \dd x,\qquad 
\Li_1(z) = - \log(1-z)\,,
\end{equation}
constitute simple examples of MPLs that many of you may have seen. 

As it turns out, MPLs are sufficient to express all one-loop Feynman integrals, as well as certain amplitudes and (the energy integrals of) conformally-coupled cosmological correlators to all loop orders. The recognition that these functions constitute the ``correct'' objects to study in the context of the simplest classes of Feynman integrals led to the so-called ``polylog revolution'' that helped push the state of the art for many phenomenologically-relevant scattering amplitudes from one to two loops. This rapid progress was fueled by the fact that the properties of MPLs had already been studied in depth by mathematicians, making it possible for physicists to quickly develop practical tools for working with these functions, including:
\begin{enumerate}
\item[$\bullet$] sophisticated methods for evaluating Feynman integrals via direct integration or using differential equations,
\item[$\bullet$] systematic algorithms for organizing MPLs into a basis,
\item[$\bullet$] efficient routines for evaluating MPLs numerically,
\item[$\bullet$] and techniques for making the analytic structure of MPLs completely transparent. 
\end{enumerate}
We will touch upon only some of these ideas in this lecture; for a more comprehensive review, see~\cite{Weinzierl:2022eaz}. 

\subsection{The Shuffle and Stuffle Algebra}
One of the key properties of MPLs is that they obey a \emph{shuffle algebra}. This can be shown by simply breaking up the unordered integration variables into different orderings. For example, consider
\begin{equation}
G_a(z) G_b(z) = \int_0^z \frac{\dd x_1}{x_1 - a} \int_0^z \frac{\dd x_2}{x_2 - b}.
\end{equation}
While the two integration variables $x_1$ and $x_2$ are unordered, we can triangulate the integration region into two regions where they are:  
\begin{equation}
\begin{tikzpicture}[scale=0.5, baseline, shift={(0,-3)}]
\draw[->, thick] (-2, 0) -- (8, 0);
\draw[->, thick] (0, -2) -- (0,8);
\draw (0, 5) -- (5,5) -- (5, 0);
\node at (8, -0.65) {$x_1$};
\node at (0.65, 8) {$x_2$};
\node at (5, -0.65) {$z$};
\node at (-0.65, 5) {$z$};
\fill[pattern=north east lines] (0,0) rectangle (5,5);
\end{tikzpicture}
\longrightarrow
\begin{tikzpicture}[scale=0.5, baseline, shift={(0,-3)}]
\draw[->, thick] (-2, 0) -- (8, 0);
\draw[->, thick] (0, -2) -- (0,8);
\draw (0, 5) -- (5,5) -- (5, 0);

\node at (8, -0.65) {$x_1$};
\node at (0.65, 8) {$x_2$};
\node at (5, -0.65) {$z$};
\node at (-0.65, 5) {$z$};
\fill[pattern=north east lines, pattern color=brown] 
    (0,0) -- (0,5) -- (5,5) -- cycle;
\fill[pattern=north east lines, pattern color=blue] 
    (0,0) -- (5,5) -- (5,0) -- cycle;

\end{tikzpicture} .
\end{equation}
Algebraically, this means we split our original integration region into the parts $\{0\leq x_1 \leq x_2 \leq z\} \cup \{0 \leq x_2 \leq x_1 \leq z\}$, within each of which we can rewrite our twofold integral as an iterated integral. Namely, 
\begin{align}
G_a(z) G_b(z) &=  \int_0^z \frac{\dd x_1}{x_1 - a} \int_0^{x_1} \frac{\dd x_2}{x_2 - b} 
+ 
 		  \int_0^z \frac{\dd x_1}{x_1 - b} \int_0^{x_1} \frac{\dd x_2}{x_2 - a} \\
&= G_{a,b}(z)  + G_{b,a}(z).
\end{align}
Repeating this argument in higher dimensions one finds that
\begin{equation}
G_{a_1, \cdots, a_n}(z) G_{b_1, \cdots, b_n}(z) = \sum_{\s \in \text{sh}(\{a_i\}, \{b_i\})} G_\s(s)
\end{equation}
where $\text{sh}(\{a_i\}, \{b_i\})$ is the set of all shuffles of the $a_i$ and $b_i$ indices, namely all ways of interleaving these indices together such that the ordering of the sets $\{a_i\}$ and $\{b_i\}$ are separately maintained (like shuffling two halves of a deck of cards). 

MPLs also obey a \emph{stuffle algebra}, which encodes the identities that arise from writing MPLs as nested sums. Namely, just as one can triangulate integration regions, one can rewrite products of MPLs, which involve unordered summation indices, as a linear combination of higher-weight MPLs that involve ordered summation indices. Both shuffle and stuffle relations are automatically taken into account when we work with the \emph{symbol} of MPLs, which we now introduce.

\subsection{The Symbol}
Starting from the definition~\eqref{eq:mpl_def}, one can compute the total derivative of a generic MPL:
\begin{equation}
d G_{a_1, \cdots, a_n}(z) = \sum_{i = 1}^n G_{a_1, \cdots, \hat a_{n-i+1},\cdots, a_n}(z) d \log \left( \frac{a_{n-i+1} - a_{n-i}}{a_{n-i+1} - a_{n-i+2}} \right)
\end{equation}
where a hat over $\hat a_{n-i+1}$ indicates this index should be dropped, and we have used the shorthand notation $a_0 = z$,  $a_{n+1} = 0$. This differential can be upgraded to (recursively) define the \emph{symbol} of an MPL:
\begin{equation}
\mathcal S\big(G_{a_1, \cdots, a_n}(z) \big) = \sum_{i =1}^n \mathcal S \big( G_{a_1, \cdots, \hat a_{n-i+1}, \cdots, a_n}(z)  \big)\otimes  \left( \frac{a_{n-i+1} - a_{n-i}}{a_{n-i+1} - a_{n-i+2}} \right).
\end{equation}
Namely, the symbol is an $n$-fold tensor object that we can associate to any weight-$n$ MPL, whose entries have the distributive properties of logarithms,
\begin{align}
\cdots \otimes ab \otimes \cdots &= \cdots \otimes a \otimes \cdots + \cdots \otimes b \otimes \cdots, \\
\cdots \otimes \frac{a}{b} \otimes \cdots &= \cdots \otimes a \otimes \cdots - \cdots \otimes b \otimes \cdots\, ,
\end{align}
since they have arisen as the arguments of $d\log$s. The entries which appear in this tensor product are referred to as \emph{symbol letters}. As an example, going back to the example of classical polylogs, 
\begin{equation}
\Li_n(z) = - G_{\underbrace{0,\cdots, 0}_{n-1},1}(z) 
\end{equation}
we find the symbol
\begin{equation} \label{eq:classical_polylog_symb}
\mathcal S\big( \Li_n(z)\big) = - (1- z) \otimes \underbrace{z \otimes z \cdots \otimes z}_{n-1}.
\end{equation}
Note here that the overall minus sign cannot be absorbed into any of the symbol letters, since the entries in the symbol are implicitly wrapped in logarithms. Sometimes it is useful to keep the logarithms explicit, such as 
\begin{equation*}
\mathcal S\big( \Li_n(z)\big) = - \log(1- z) \otimes \underbrace{\log(z) \otimes \log(z) \cdots \otimes \log(z)}_{n-1},
\end{equation*}
however the symbol is usually represented as in~\eqref{eq:classical_polylog_symb}.

The symbol inherits the shuffle algebra of the MPLs, namely 
\begin{equation}
\mathcal S (F_1 F_2) = \mathcal S(F_1)  \shuffle \mathcal S(F_2),
\end{equation}
where $\shuffle$ indicates the sum over all shuffles of the entries in $\mathcal{S}(F_1)$ and $\mathcal{S}(F_2)$. For example,
\begin{align}
\mathcal S\big( \log(z) \Li_2(y) \big) &= \mathcal S(\log(z)) \shuffle \mathcal S(\Li_2(y)) \\
&= - [z] \shuffle [(1-y) \otimes y] \\
&= - z \otimes (1-y) \otimes y  - (1-y) \otimes z \otimes y - (1-y) \otimes y \otimes z\,,
\end{align}
where the symbol letter $(1-y)$ appears before the letter $y$ in every term, as per the definition of the shuffle product.

Symbols are much easier to work with than full MPLs. For instance, it can be easily checked by hand that
\begin{equation}
\mathcal S(\log(z) \log(1-z)) = \mathcal S(-\Li_2(z) - \Li_2(1-z)).
\end{equation}
Referring back to the definition of the symbol, this means that the (sequences of) derivatives of these functions are the same, modulo constants. Thus, these two functions must be the same functions when considered as iterated integrals, up to integration constants. A key result from Goncharov \cite{Goncharov:2009lql} ensures that this is indeed the case. Thus, by systematically adding terms that involve transcendental constants, any identity that holds at symbol level can be upgraded to an identity between full MPLs. For instance, if we take $z \to 0$ in the example above, the product of logs vanishes while the sum of dilogs becomes $-\zeta_2$. 
It can be checked numerically that this is the only additional contribution needed, to upgrade this identity to full function level:
\begin{equation}
\log(z) \log(1-z) = - \Li_2(z) - \Li_2(1-z) + \zeta_2.
\end{equation}
Note that in more general examples, one might also need to include weight-one constants like $i\pi$ or $\log(2)$ multiplying a function of $z$. 

\begin{figure}[t]
	\centering
	\includegraphics{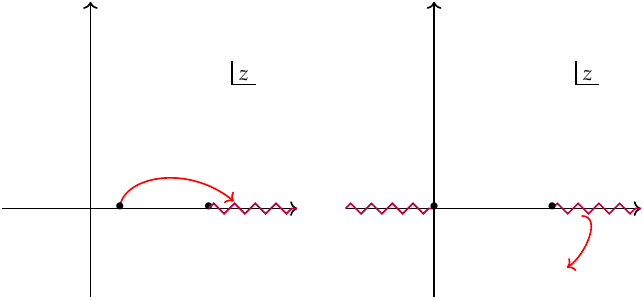}
\caption{A new branch cut appears at $z=0$ in the dilogarithm if we analytically continue around the branch point at $z=1$.}
	\label{DiLogbranch2}
\end{figure}

Now, rather than think about the symbol as encoding the derivatives of an MPL, let's think about what it tells us about an MPL's discontinuities. Consider again the dilogarithm
\begin{equation} \label{eq:dilog}
\Li_2(z) = -\int_0^z \frac{\log(1-x)}{x} d x\, ,
\end{equation}
whose symbol is given by
\begin{equation} \label{eq:dilog_symbol}
    \mathcal{S}\left( \Li_2(z) \right) = -(1-z) \otimes z \, .
\end{equation}
Where are the branch points of $\Li_2(z)$? 
\begin{figure}[t]
	\centering
	\includegraphics{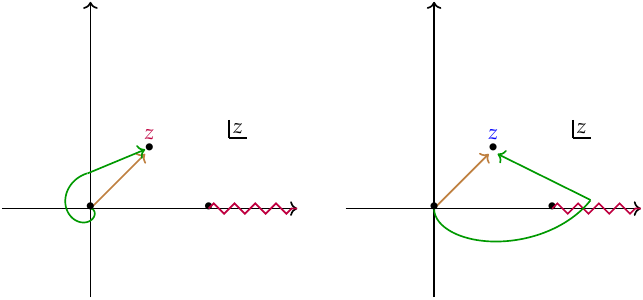}
\caption{The integration contours for the dilogarithm before we have analytically continued (in red), and after we have analytically continued around either $z=0$ or $z=1$ (in green).}
	\label{Li2branchcut}
\end{figure}
We can answer this question at two levels. First, the symbol allows us to easily identify the placement of all of $\Li_2(z)$'s branch points in the $z$ plane; these can be read directly off of the letters that appear, which record where the MPL has logarithmic singularities. Inspecting~\eqref{eq:dilog_symbol}, we see that $\Li_2(z)$ has branch points at $z=0$ and $z=1$. However, are these branch points both visible on the principal Riemann sheet? We can probe this by analytically continuing $\Li_2(z)$ around these two points. 
\begin{enumerate}
\item[$(0)$] If we analytically continue $z$ around a path that encircles 0 (but not 1), the value of $\Li_2(z)$ remains unchanged. We can see this by thinking back to the integral representation in~\eqref{eq:dilog}; in particular, we note that we do not encircle the branch point that appears in the integrand $\log(1-x)/x$ when we carry out this analytic continuation (since the analytic continuation of the contour does not force $x$ to go around $x=1$). However, since there is a simple pole at $x=0$, the analytically continued contour is not in the same homotopy class as the original one. Rather, we can split up the new integration contour into the original contour plus a contour that encircles $x=0$:
\begin{align}
    -\int_0^z \frac{\log(1-x)}{x} dx \ \ \xrightarrow[\text{continuation}]{\text{analytic}} \ \ -\int_0^z \frac{\log(1-x)}{x} dx -\int_{\circlearrowleft_{x=0}} \frac{\log(1-x)}{x} dx  \, .
\end{align}
The second integral simply computes the residue of the integrand at $x=0$, but this evaluates to zero due to the logarithm in the numerator. Thus, we see the value of this integral remains unchanged. This signals that the branch point at $z=0$ is not visible on the principal Riemann sheet.
\item[$(1)$] If we instead analytically continue $z$ around 1 (but not 0), the branch cut in the integrand $\log(1-x)/x$ will now play a role. Similar to before, we split up the analytically continued contour into the original contour plus a contour that encircles $x=1$:
\begin{align}
    -\int_0^z \frac{\log(1-x)}{x} dx \ \ \xrightarrow[\text{continuation}]{\text{analytic}} \ \ -\int_0^z \frac{\log(1-x)}{x} dx -\int_{\circlearrowleft_{x=1}} \frac{\log(1-x)}{x} dx  \, .
\end{align}
Since the only special points in the integrand are $x=0$ and $x=1$, we can deform the second contour into a line that goes directly from $z$ to $1$, followed by a small circle around $1$, and then the same line as before traversed in the opposite direction, from $1$ back to $z$. The integral along the small circle will vanish, since the residue at $x=1$ is zero. However, the integrations from $z$ to $1$ and back will not cancel, as the integrand will pick up an additional contribution $2\pi i/x$ when we traverse a small circle around $x=1$ (which we assume we have done counterclockwise). This means we will get a new contribution 
\begin{equation} \label{eq:disc_dilog_1}
-\int_1^z \frac{2 \pi i}{x} dx = - 2 \pi i \log(z).
\end{equation}
Thus, we see that the branch point at $z=1$ is indeed present on the principal Riemann sheet.
\end{enumerate}
Having seen that the branch point at $z=0$ is not visible on the principal Riemann sheet, we might wonder where it is. However, we can see the answer to this question quite clearly in~\eqref{eq:disc_dilog_1}. The discontinuity 
\begin{equation}
    \textrm{Disc}_{z-1}\Li_2(z) = - 2 \pi i \log(z)
\end{equation} 
itself has a branch point at $z=0$. This is illustrated in Figure~\ref{DiLogbranch2}. Thus, the branch point at $z=0$ is simply hidden on the second Riemann sheet.  

\FloatBarrier

Looking back at the dilog's symbol~\eqref{eq:dilog_symbol}, we see this information about the Riemann sheet structure is also transparently displayed for us. Instead of reading the symbol from right to left (which we do to understand the derivatives of MPLs), we simply read the symbol from left to right. In this case, we see we initially just have a nonzero discontinuity starting at $z=1$, since $(1-z)$ is the only letter that appears in the first entry of the symbol. However, if we compute a discontinuity with respect to this branch cut, this effectively replaces the letter $(1-z)$ in the first entry by a factor of $2 \pi i$: 
\begin{equation}
    -(1-z) \otimes z \xrightarrow{\textrm{Disc}_{z-1}} 2 \pi i \otimes z 
\end{equation}
This newly uncovers a branch point at $z=0$. This lesson generalizes to symbols that involve more than one term. For instance, computing the symbol of a generic weight-two MPL, we have
\begin{align}
\mathcal S\big( G_{a,b}(z)\big) &= a \otimes b - a \otimes (b-a) + b \otimes (b-a) - b \otimes (z - a) - (z- a) \otimes b\\\notag
&+ (z-a)\otimes (b-a) - (z-b) \otimes(b-a) + (z -b) \otimes (z-a).
\end{align}
Each term in this symbol encodes a nonzero double discontinuity; for instance the last term tells us that 
\begin{equation}
   \text{Disc}_{z-a}(\text{Disc}_{z-b}( G_{a,b}(z))) = (2 \pi i)^2\, .
\end{equation}
Clearly, reading this information off of the symbol is much easier than computing it via explicit analytic continuation.

\subsection{Analytic Constraints on the Symbol}

Given the direct connection between the symbols of MPLs and their logarithmic discontinuities, we can now easily see how the types of analytic properties we derived in the previous lectures can be used to constrain the functions that amplitudes, correlators, and related quantities evaluate to. Namely, if we can convince ourselves that the quantity we are studying evaluates to an MPL (which we can often do, using methods we will mention shortly), then $\alpha$-positivity, hierarchical, and Steinmann-type constraints can be used to directly constrain its symbol:
\begin{enumerate}
\item[$\bullet$] Only $\alpha$-positive singularities can appear directly as letters in the first entry of the symbol.
\item[$\bullet$] Hierarchical constraints become restrictions on which ordered pairs of symbol letters can appear in symbol terms (with any number of other letters between them). 
\item[$\bullet$] Steinmann-type constraints restrict which letters can appear directly next to each other in terms of the symbol. For this reason, this class of constraints are often referred to as \emph{adjacency constraints}.
\end{enumerate}
Working out the constraints implied by $\alpha$ positivity, the hierarchical principle, and the (extended) Steinmann relations becomes a bit more involved once algebraic singularities begin to appear; however, these complications are under good theoretical control (see for instance~\cite{Hannesdottir:2024hke}). 

Because of how transparently the symbol encodes the analytic structure of iterated integrals, bootstrap calculations are often carried out at symbol level first, and then upgraded to full function level only once the symbol is uniquely determined. (Note that this statement isn't restricted to MPLs, as bootstrap calculations have also been carried out for integrals involving elliptic curves~\cite{Morales:2022csr}.) This latter step involves determining the contributions that are proportional to constants such as $\zeta_3$, which are annihilated by the symbol map. However, assuming that one has a reasonable guess for the constants that can appear, this step is also doable. We will see an example of this in lecture \ref{sec:bootstrap}.

\subsection{Elliptics and Beyond}
Of course, as indicated earlier, it is not always the case that Feynman integrals can be evaluated in terms of MPLs. For instance, when performing a logarithmic integration, it is possible to generate a new square root:
\begin{equation}
\int_0^\infty \frac{d\a}{\a^2 + 2 f \a + g} = \frac{1}{2\sqrt{f^2 - g}} \log \left(  \frac{f - \sqrt{f^2 - g}}{f + \sqrt{f^2 - g}} \right)\, .
\end{equation}
If, moreover, the quantity $f^2 - g$ depends on further integration variables, it may not be possible to integrate them in terms of MPLs. More specifically, one of the two things may happen:
\begin{enumerate}
\item[$(i)$] either a change of variables may be found to rationalize the root, so that we can carry out another integration algebraically or (poly)logarithmically,
\item[$(ii)$] or, if the polynomial is degree three or higher in all remaining integration variables and the root cannot be rationalized, we have hit an obstruction to integration via MPLs. 
\end{enumerate}
The simplest example of the type of obstruction described in $(ii)$ appears in the massive sunrise integral in $D=2$:
\begin{equation}
\begin{tikzpicture}[baseline]
\draw (-2,0) -- (2,0);
\draw (0,0) circle (1);
\node at (0,1.25) {$m_1$};
\node at (0,0.25) {$m_2$};
\node at (0,-1.25) {$m_3$};
\end{tikzpicture}
\end{equation}
If we start from the Feynman parameter representation and evaluate two of the integrations, we are left with a single integral that takes the schematic form 
\begin{equation}
    \sim \int_0^\infty \frac{\log(\dots)}{\sqrt{\alpha^4 + \cdots}} d \alpha \,.
\end{equation}
Notice that the remaining integrand has no simple poles; rather, the square root in the denominator encodes an elliptic curve via the equation $y^2 = \alpha^4 + \dots$~\cite{Bloch:2013tra}. This means that we can only evaluate the massive sunrise integral in terms of iterated integrals if we enlarge our space of functions to include integration kernels that depend on this elliptic curve.

In fact, if we upgrade the sunrise to an $L$-loop banana integral by adding additional massive propagators between the same pair of vertices, we instead end up encountering a square root that is irreducibly quartic (or cubic, with a root at infinity) with respect to $L-1$ integration variables:
\begin{equation} \label{eq:banana}
\begin{tikzpicture}[baseline]
\draw (-2,0) -- (-1,0);
\draw (2,0) -- (1,0);
\draw (0,0) circle (1);
\draw (-1,0) to[out = 60, in=120] (1,0);
\draw (-1,0) to[out = -60, in=-120] (1,0);

\node at (0,0) {$\vdots$};
\end{tikzpicture} \sim \int \dd\a_1 \cdots \dd\a_{L-1} \frac{\log(\cdots)}{\sqrt{f_{L-1}^{CY}}}\, .
\end{equation}
It turns out that the equation $y^2 = f_{L-1}^{CY}$ defines an $(L-1)$-dimensional Calabi-Yau (CY) manifold, indicating that periods of these CY manifolds must appear in the special functions the banana integrals evaluate to. This means the complexity of this class of diagrams grows in an unbounded way with the loop order! 

In the banana integrals, obstructions to integration via MPLs appear after just a single logarithmic integration has been carried out. More generally, this need not be the case; we can also encounter such obstructions after having evaluated many integrations in terms of MPLs. Ideally, then, we would like to be able to predict whether or not such obstructions will arise \emph{before} we start integrating. This can be done by computing \emph{leading singularities}, which amounts to replacing the physical integration contour with residue contours. The intuition is that these residues simulate polylogarithmic integrations (or track the sequences of logarithmic discontinuities these integrations encode); thus, if we compute all possible sequences of residues, we should encounter the same obstructions that would have appeared if we had integrated over the original contour. This gives us a way to predict the types of special functions that we expect a given Feynman integral to evaluate to. In particular, if we can evaluate all integrations via arbitrary sequences of residues, we conclude that the corresponding integral can be evaluated in terms of MPLs.

Using leading singularities, one can show that the complexity of the integrals we encounter in massless diagrams also grows rapidly with the loop order. For instance, like the banana integrals, the train-track diagrams
\begin{equation}
\begin{tikzpicture}[baseline={(0,0.4)}]
\draw (-4,1) -- (4,1);
\draw (-4,0) -- (4,0);
\draw (-3.5, 1.25) -- (-3.5, -0.25);
\draw (-2.5, 1.25) -- (-2.5, -0.25);
\draw (-1.5, 1.25) -- (-1.5, -0.25);
\node at (0,0.5) {$\cdots$};
\draw (3.5, 1.25) -- (3.5, -0.25);
\draw (2.5, 1.25) -- (2.5, -0.25);
\draw (1.5, 1.25) -- (1.5, -0.25);
\end{tikzpicture}
\end{equation}
in $D=4$ give rise to integrals over $(L-1)$-dimensional CY manifolds at $L-$loops~\cite{Bourjaily:2018ycu}.  This growth in complexity becomes even more drastic if we go to the non-planar sector; for instance, the massless tardigrade integrals
\begin{equation}
\begin{tikzpicture}[baseline={(0,0)}]
\draw (-2, 0.5) -- (0,0.5);
\draw (1, 0.5) -- (2,0.5);
\draw (-2, -0.5) -- (0, -0.5);
\draw (1, -0.5) -- (2, -0.5);
\draw (2, 0.5) -- (3,0);
\draw (2, -0.5) -- (3,0);
\draw (-2, 0.5) -- (-1, -0.5);
\draw (-1, 0.5) -- (0, -0.5);
\draw (1, 0.5) -- (2, -0.5);
\draw (-2, -0.5) -- (-1.55, -0.05);
\draw (-1.45,  0.05) -- (-1, 0.5);
\draw (-1, -0.5) -- (-0.55, -0.05);
\draw (-0.45,  0.05) -- (0, 0.5);
\draw (1, -0.5) -- (1.45, -0.05);
\draw (1.55,  0.05) -- (2, 0.5);
\draw[dashed] (0, 0.5) -- (1, -0.5);
\draw[dashed] (0, 0.5) -- (1, 0.5);
\draw[dashed] (0, -0.5) -- (1, -0.5);
\draw[dashed] (0, -0.5) -- (1, 0.5);
\draw (-2, 0.5) -- (-2.45, 0.5+0.4);
\draw (-2, 0.5) -- (-2 -0.45, 0.5-0.4);
\draw (-2, -0.5) -- (-2.45, -0.5+0.4);
\draw (-2, -0.5) -- (-2 -0.45, -0.5-0.4);
\draw (2, 0.5) -- (2.5, 0.5+0.4);
\draw (2, -0.5) -- (2.5, -0.5-0.4);
\draw (3, 0) -- (3+0.5, 0.4);
\draw (3, 0) -- (3+0.5, -0.4);
\end{tikzpicture} 
\end{equation}
give rise to integrals over $(2L-2)$-dimensional CY manifolds~\cite{Bourjaily:2018yfy}. This means we already encounter a K3 integral at two loops, in the diagram
\begin{equation}
    \begin{tikzpicture}[baseline]
\draw (-1.4,.2) -- (-1,0);
\draw (-1.4,-.2) -- (-1,0);
\draw (-1,0) -- (0,1);
\draw (-1,0) -- (0,-1);
\draw (0,-1) -- (1,0);
\draw (0,1) -- (1,0);
\draw (1,0) -- (1.4,0.2);
\draw (1,0) -- (1.4,-0.2);
\draw (0,1) -- (2.5,0);
\draw (0,-1) -- (2.5,0);
\draw (2.5,0) -- (2.9,-0.2);
\draw (2.5,0) -- (2.9,0.2);
\draw (0,1) -- (-0.25,1.25);
\draw (0,1) -- (0.25,1.25);
\draw (0,-1) -- (-0.25,-1.25);
\draw (0,-1) -- (0.25,-1.25);
\end{tikzpicture} 
\end{equation}
Note that this diagram contributes to many simple quantum field theories, such as massless $\phi^4$ theory.

At this point, it would be natural to wonder whether only CY manifolds appear when we consider Feynman integrals. However, the answer is no. For instance, the massive necklace diagrams 
\begin{equation}
\begin{tikzpicture}
\draw (1*360/7+90: 2cm) to[bend right=60] (1*360/7+360/7+90: 2cm) ;
\draw (2*360/7+90: 2cm) to[bend right=37] (2*360/7-360/7+90: 2cm) ;
\draw (2*360/7+90: 2cm) to[bend right=60] (2*360/7+360/7+90: 2cm) ;
\draw (3*360/7+90: 2cm) to[bend right=37] (3*360/7-360/7+90: 2cm) ;
\draw (4*360/7+90: 2cm) to[bend right=60] (4*360/7+360/7+90: 2cm) ;
\draw (5*360/7+90: 2cm) to[bend right=37] (5*360/7-360/7+90: 2cm) ;
\draw (5*360/7+90: 2cm) to[bend right=60] (5*360/7+360/7+90: 2cm) ;
\draw (6*360/7+90: 2cm) to[bend right=37] (6*360/7-360/7+90: 2cm) ;
\draw (6*360/7+90: 2cm) to[bend right=60] (6*360/7+360/7+90: 2cm) ;
\draw (7*360/7+90: 2cm) to[bend right=37] (7*360/7-360/7+90: 2cm) ;
\draw (7*360/7+90: 2cm) to[bend right=60] (7*360/7+360/7+90: 2cm) ;
\draw (1*360/7+90: 2cm) to[bend right=37] (1*360/7-360/7+90: 2cm) ;
\node (m1) at (-0.52066, 1.08) {$m_6$};
\node (m2) at (-1.15, 0.25) {$m_4$};
\node (m3) at (-0.93, -0.72) {$m_2$};
\node (m5) at (0.882375, -0.72) {$m_{12}$};
\node (m6) at (1.043, 0.25) {$m_{10}$};
\node (m7) at (0.52066, 1.08) {$m_8$};
\node (mm1) at (-1.08471, 2.25242) {$m_5$};
\node (mm2) at (-2.66, 0.611932) {$m_3$};
\node (mm3) at (-2.08788, -1.6) {$m_1$};
\node (mm5) at (2.13049, -1.6) {$m_{11}$};
\node (mm6) at (2.6, 0.623058) {$m_9$};
\node (mm7) at (1.08471, 2.25242) {$m_7$};
\node (mm4) at (0,-2.) {\large $\dots$};
\end{tikzpicture}
\end{equation}
provide a counterexample in $D=2$, as they can be shown to give rise to hyperelliptic curves of genus $L-1$ (which are not CY for $L>3$) at $L$ loops~\cite{Abreu:2024jde}. While these are just vacuum diagrams, note that this places a lower bound on the types of special functions that must appear in diagrams that reduce to these topologies in soft limits. 

As a final comment, let us highlight that leading singularities also teach us about the rational and algebraic prefactors that appear in Feynman integrals. For instance, let's imagine that we have managed to evaluate all of the integrations in some Feynman integral via a sequence of residues. The resulting function of external kinematic variables is precisely what we should be left with if we were to compute some (as yet unspecified) maximal sequence of logarithmic discontinuities. That is, since each logarithmic discontinuity simply generates a factor of $2 \pi i$, any maximal (nonvanishing) sequence of discontinuities of a pure weight-$n$ MPL will simply evaluate to $(2 \pi i)^n$. Thus, when constructing the ansatz for a given Feynman integral, we ought to include the leading singularities of this integral as the prefactors of whatever transcendental functions we expect will arise, in order to incorporate this expected kinematic dependence. 


\section{Bootstrap Methods}
\label{sec:bootstrap}
In this final lecture, we will see how all the ideas we have discussed come together to allow us to bootstrap the functional form of integrals directly. The method we will follow can be sketched as follows~\cite{Hannesdottir:2024hke}:
\begin{enumerate}
    \item[$(i)$] Identify the types of special functions that can appear (and the rational or algebraic prefactors they are expected to multiply), by computing leading singularities.
    \item[$(ii)$] Identify the kinematic locations where singularities can appear, and construct symbol letters (or more generally, integration kernels) that are consistent with this set of singularities (including the type of singularity that will arise at each location, if known).
    \item[$(iii)$] Derive constraints on the sequences of discontinuities that can appear in the final result ($\alpha$-positivity, hierarchical constraints, adjacency relations).
\end{enumerate}

\noindent Combining our knowledge from steps $(i)$-$(iii)$ with our expectation for the maximum transcendental weight that can appear~\cite{Hannesdottir:2021kpd}, we can now construct the (finite) space of functions that we believe our Feynman integral will evaluate to. We then usually need to carry out one further step to determine the unique function within this space that corresponds to our target integral:
\begin{enumerate}[resume]
\item[$(iv)$] Fix any remaining freedom matching our ansatz to the expansion of the target integral around singular kinematic points.
\end{enumerate}
We now illustrate this method in some simple examples.

\subsection{The Massive Bubble Integral (again)}
\begin{figure}[t]
	\centering
	\includegraphics{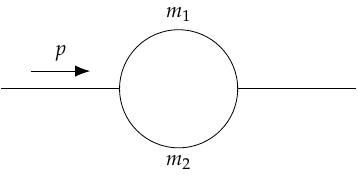}
    \caption{The massive bubble integral.}
	\label{fig:bubble}
\end{figure}

Let us consider again the massive bubble integral in $D=2$, which we have drawn in Figure \ref{fig:bubble}. We will now show we can bootstrap this integral just from knowledge of its singularities and discontinuities. 

\FloatBarrier

\renewcommand{\thesubsubsection}{(i)}
\subsubsection{Special functions and algebraic prefactors}

As a first step, we want to probe whether there exists an obstruction to evaluating this integral in terms of MPLs. As described in the last section, we do this by computing all possible sequences of residues of the integrand. Recall that we computed these residues back in section~\ref{sec:massive_bubble}, and found they were all proportional to
\begin{align}\label{eq:bubble-onefold}
        \sim \frac{1}{\sqrt{s-r_+}\sqrt{s-r_-}}\, ,
\end{align}
where we have introduced the notation $r_\pm \coloneqq (m_1\pm m_2)^2$. This tells us that it should be possible to do all the integrations either algebraically or (poly)logarithmically. In addition, since there is a unique leading singularity, it tells us that all of the iterated integrals in the final answer should multiply this algebraic factor.

\renewcommand{\thesubsubsection}{(ii)}
\subsubsection{Singularities and Letters}    

Next, we work out where singularities can arise in the final result. At one loop, this can be done by solving the Landau equations. At higher loops, finding all singularities generally requires using blowups, so it is better to use one of the tools that implements an alternate method for efficiently predicting a (possibly overcomplete) set of singular loci (see for instance~\cite{Panzer:2014caa} and ~\cite{Correia:2025wtb}). The possible singular loci for the massive bubble in $D=2$ are shown in Table \ref{tab:landau-bubble}.\footnote{For explicit examples of solving the Landau equations directly, see the exercises.}
\begin{table}[]
        \centering
        \begin{tabular}{c|c|c|c}
            singularity & which propagators on shell? & $\alpha$-positive? & nature of singularity\\
            \hline
             $m_1^2=0$ & 1 & yes & $\text{logarithmic}$ \\
             $m_2^2=0$ & 2 & yes & $\text{logarithmic}$ \\
             $s=r_+$ & 1,2 & yes & $\text{square root}$ \\
             $s=r_-$ & 1,2 & no & $\text{square root}$
        \end{tabular}
        \caption{The possible singularities  for the massive bubble that we get by solving the Landau equations.}
        \label{tab:landau-bubble}
\end{table}

The nature of the singularities that appear in the massive bubble can also be deduced using the formula derived by Landau, which we quoted in~\eqref{eq:singularity_nature} and recall here:
    \begin{align} \label{eq:asymptotics_sec_5}
        I(\varphi\to0)=
        \begin{cases}
\varphi^{\gamma}\log\varphi & \text{if $\gamma\in \mathbb{Z}$ and $\gamma\geq0$} \\
\varphi^{\gamma} & \text{otherwise}
\end{cases}
    \end{align}
where $\gamma=(DL-E-1)/2$. For the bubble, we have $D=2$ and $L=1$, while $E$ represents the number of propagators that are placed on shell by the corresponding solution to the Landau equations. Thus, referring to Table~\ref{tab:landau-bubble}, we see that $E=1$ for the singularities at $m_i^2=0$, and $E=2$  for the singularities at $s = r_\pm$. This implies that $\gamma=0$ for the mass singularities, so we expect them to be logarithmic, while $\gamma=-\frac12$ for the threshold and pseudo-threshold, so they should be algebraic.

Next, we want to construct all possible symbol letters that encode these (and only these) singularities. Singularities that are of logarithmic type can appear as letters by themselves. Thus, the two squared masses, 
    \begin{align} \label{eq:rational_alphabet}
        \log(m_1^2) \,, \quad \log(m_2^2)\, ,
    \end{align}
constitute candidate letters.

What about algebraic symbol letters? In practice, we can construct candidate algebraic letters using the algorithm introduced in~\cite{Heller:2019gkq}. We begin by assuming these letters take the form \begin{equation} \label{eq:letter_form}
    \log(A\pm\sqrt{B}) \, ,
\end{equation}
since this resembles what has been observed in previous calculations of Feynman integrals. Our goal is then to determine what polynomials (of $m_1$, $m_2$, and $s$) can be plugged in for $A$ and $B$. The first thing we can input as a constraint is that the zeros of $B$ must correspond to the loci where square root singularities can appear. This means that 
    \begin{align} \label{eq:B_ansatz}
        B = \prod_j R_j,
    \end{align}
where the product over $j$ ranges over all possible subsets of square root branch points.\footnote{Note that we don't consider different powers of each factor $R_j$ because we are assuming $B$ is a polynomial, meaning any powers should be integer. However, if $R_j$ were to appear to a power $p>1$, we would just pull a factor of $R_j^{\lfloor p/2 \rfloor}$ out of the square root (and by redefining $A$, out of the letter altogether).} Referring back to Table~\ref{tab:landau-bubble}, we have $R_j \in \{ s- r_+, s- r_-\}$ for the bubble integral. Next, we note that letters taking the form~\eqref{eq:letter_form} can give rise to logarithmic branch points where the product $(A+\sqrt{B})(A-\sqrt{B}) = A^2-B=0$ vanishes. This implies that we must have
\begin{align}
        A^2-B=c\prod_i L_i^{n_i},
\end{align}
where each of the $L_i$ represents one of  the possible logarithmic branch points, the $n_i$ are undetermined integer powers, and $c$ is an undetermined number. For the bubble, we have that $L_i\in\{m_1^2,m_2^2\}$. Putting this constraint together with~\eqref{eq:B_ansatz}, we can construct an ansatz for $A^2$:
\begin{align}
        A^2 = c \prod_i L_i^{n_i} + \prod_j R_j \, .
\end{align}
Using the fact that the products draw the $L_i$ and $R_j$ factors from finite sets, we can range over integer values for the $n_i$, and ask whether there exists a number $c$ that makes the right hand side a perfect square. If, for a given choice of factors $L_i$, $R_j$ and integers $n_i$, no such $c$ exists, then no permissible letter of the form~\eqref{eq:letter_form} for these choices exists. Note that, in practice, we do not need to consider arbitrarily high powers of $n_i$, since the polynomial degree of $\prod_j R_j$ is bounded, and thus there is only a chance of forming a perfect square if the powers $n_i$ are not too large.

One typically finds that there exist only a few solutions to the above constraints. In the bubble example, we find just the pair
    \begin{align}
        \log(s-r_+ + \sqrt{s-r_+}\sqrt{s-r_-}), \quad \log(s-r_+ -\sqrt{s-r_+}\sqrt{s-r_-}).
    \end{align}
One can check that the product of these two symbol letters is proportional to $m_1^2m_2^2$, so this only actually corresponds to one new (multiplicatively independent) letter. Without loss of generality, we thus define a new letter
\begin{equation} \label{eq:algebraic_letter}
    \log(R) = \log\left(\frac{s-r_+-\sqrt{s-r_+}\sqrt{s-r_-}}{s-r_+ + \sqrt{s-r_+}\sqrt{s-r_-}}\right)\, ,
\end{equation}
which we append to those in~\eqref{eq:rational_alphabet} to form our candidate symbol alphabet.

\renewcommand{\thesubsubsection}{(iii)}
\subsubsection{Analytic Constraints} 

Synthesizing our knowledge from steps $(i)$ and $(ii)$, we can formulate an ansatz for the symbol of the bubble integral:
    \begin{align} \label{eq:ansatz_iii}
        \mathcal{S}(I_G) = \frac{1}{\sqrt{s-r_+}\sqrt{s-r_-}} \left(\substack{\text{A polylogarithmic symbol constructed} \\ \text{from $\log(m_1)$, $\log(m_2)$, $\log(R)$}} \right).
    \end{align}
We will next restrict what sequences of letters can appear in this symbol, using the hierarchical principle. Since the $m_i^2=0$ singularity can only be reached when propagator $i$ is on shell (and not when both propagators are on shell), we immediately have that
    \begin{align}
        &\Disc_{m_1^2}\cdots\Disc_{m_2^2}\cdots(I_G)=0,\\
        &\Disc_{m_2^2}\cdots\Disc_{m_1^2}\cdots(I_G)=0.
    \end{align}
This implies that we cannot have any symbol terms of the form
    \begin{align}
        &\cancel{\color{black} \cdots \otimes m_1^2\otimes\cdots\otimes m_2^2\otimes\cdots }\\
        &\cancel{\color{black} \cdots \otimes m_2^2\otimes\cdots\otimes m_1^2\otimes\cdots }
    \end{align}
In fact, $R$ also encodes logarithmic branch cuts that start where $m_1=0$ and $m_2=0$. This can be checked by series expanding~\eqref{eq:algebraic_letter} around $m_i \to 0$. This means that no symbol term can contain both $R$ and $m_1^2$ (and similarly for $m_2^2$), or even the letter $R$ twice. Hence, we get three more restrictions:
    \begin{align}
        &\cancel{\color{black} \cdots \otimes R\otimes\cdots\otimes m_i^2\otimes\cdots }\\
        &\cancel{\color{black} \cdots \otimes m_i^2\otimes\cdots\otimes R\otimes\cdots }\\
        &\cancel{\color{black} \cdots \otimes R\otimes\cdots\otimes R\otimes\cdots }
    \end{align}
    This leaves only three possible types of symbol terms in our ansatz:
    \begin{enumerate}[label=\roman*)]
        \item[$\bullet$] $\log(R)$ by itself,
        \item[$\bullet$] $m_1^2\otimes\cdots \otimes m_1^2$ of any length,
        \item[$\bullet$] $m_2^2\otimes\cdots\otimes m_2^2$ of any length.
    \end{enumerate}
Explicitly, we adopt the following parametrization of~\eqref{eq:ansatz_iii}:
\begin{align} \label{eq:ansatz_iii_final}
        \mathcal{S}(I_G) = \frac{1}{\sqrt{s-r_+}\sqrt{s-r_-}} \left( c_0 + c_1\log(R) + \sum_{i=1}^\infty a_i \log^i(m_1^2) + \sum_{i=1}^\infty b_i \log^i(m_2^2) \right) \, ,
    \end{align}
where we expect $\{c_0, c_1, a_i, b_i\}$ to all be rational numbers (since we have already captured the expected kinematic dependence for what multiplies the symbol).

Next, we leverage the fact that the singularity at $s=r_-$ is not $\alpha$-positive. This means there should be no branch point there on the principal Riemann sheet. To impose this requirement on our ansatz, we require that it remains the same if we analytically continue once around this branch point:
\begin{equation}
    \mathcal{S}(I_G) - \mathcal{S}(I_G|_{\sqrt{s-r_-} \to - \sqrt{s-r_-}}) = 0 \, .
\end{equation}
Since the prefactor changes sign under this analytic continuation while $\log(m_1^2)$ and $\log(m_2^2)$ remain invariant, this constraint implies that $c_0=a_i=b_i=0$. (Since $\log(R) \to - \log(R)$, no constraint is placed on $c_1$.) This leaves us with the much smaller ansatz
\begin{align} \label{eq:ansatz_iii_final_final}
        \mathcal{S}(I_G) = c_1 \frac{\log(R)}{\sqrt{s-r_+}\sqrt{s-r_-}}.
\end{align}

\renewcommand{\thesubsubsection}{(iv)}
\subsubsection{Singular Limits}

As a cross-check, we now compare the expansion of~\eqref{eq:ansatz_iii_final_final} in singular limits to the expected behavior of the bubble integral. First, we know from Landau's formula \eqref{eq:asymptotics_sec_5} that this integral should behave as $\sim \log m_i^2$ as $m_i^2\to 0$. This is indeed how our ansatz behaves. (Note that if any of the coefficients $a_i$ or $b_i$ in~\eqref{eq:ansatz_iii_final} had been found to be nonzero for $i > 1$, they would have spoiled this behavior.)

Now let's check the behavior of our ansatz near the threshold singularity $s \to r_+$. Here, we notice a problem. Namely, our ansatz is symmetric with respect to swapping $r_+ \leftrightarrow r_-$, and has been constructed to have no branch point on the principal sheet at $s = r_-$. Yet we know that the bubble integral has a nonzero discontinuity at $s = r_+$. With a little thought, we can solve this problem by dividing both the numerator and denominator of $R$ by $\sqrt{s-r_+}$:
    \begin{align}
        R\to R'\coloneqq \frac{\sqrt{s-r_+}-\sqrt{s-r_-}}{\sqrt{s-r_+}+\sqrt{s-r_-}}\, .
    \end{align}
Now, we can check that we indeed have a nonzero discontinuity at $s = r_+$, while there is no discontinuity at $s = r_-$ on principal sheet. Thus, up to a proportionality factor, we have deduced that
    \begin{align}
        I_{\bubble}^{D=2}\propto \frac{1}{\sqrt{s-r_+}\sqrt{s-r_-}}\log(R')
    \end{align}
purely from knowledge of the analytic structure that this integral ought to exhibit.

\subsection{Outlook}\label{sec:antipodal}

While there is no doubt that the bubble integral is an exceptionally simple example, these types of bootstrap methods remain in their infancy. Thus, rapid developments are expected that will help fuel advances in coming years; moreover, these bootstrap methods are ripe for use in contexts beyond collider physics, such as cosmology. For now, we just highlight that further examples can be found in \cite{Hannesdottir:2024hke,Vergu:2025mag}. In particular, it is worth looking at the example of the bubble in $D=3$, to see how different Feynman integrals can look in different numbers of dimensions (in this case, due to the fact that mass discontinuities become square roots, while the threshold and pseudo-threshold discontinuities become logarithmic). 

\begin{table}
    \centering
    \begin{tabular}{c|r}
         $L$ & number of terms  \\
         \hline
         1& 6\\
         2 & 12\\
         3 & 636\\
         4 & 11,208\\
         5 & 263,880\\
         6 & 4,916,466\\
         7 & 92,954,568\\
         8 & 1,671,656,292
    \end{tabular}
    \caption{The number of terms in the symbol of the three-particle form factor, which was bootstrapped in~\cite{Dixon:2020bbt,Dixon:2022rse}.}
    \label{tab:form-factor-symbol}
\end{table}

As a final teaser, let us also highlight how perturbative bootstrap methods have led to surprising developments in other contexts, namely in planar $\mathcal{N}=4$ supersymmetric Yang-Mills theory (where, as we already saw in section~\ref{sec:bootstrap_motivation}, they have provided us a wealth of amplitude data). The complexity of the results that have been bootstrapped in this theory cannot be overstated; for instance, see Table~\ref{tab:form-factor-symbol} for a summary of how many terms appear in the symbol of the three-particle form factor computed in~\cite{Dixon:2020bbt,Dixon:2022rse}. 

\begin{figure}[t]
	\centering
	\includegraphics{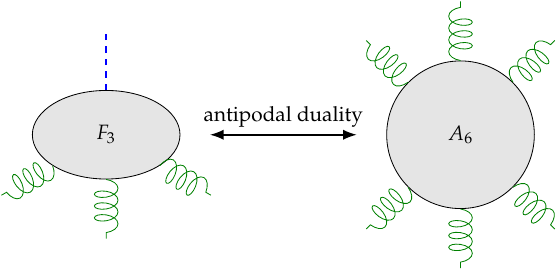}
    \caption{The antipodal duality between certain three-particle form factor and six-particle amplitudes in planar $\mathcal{N}=4$ supersymmetric Yang-Mills theory.}
	\label{fig:antipodal}
\end{figure}

Hidden in this data is an amazing plethora of surprises. Here, we will just mention the most striking of these: the discovery of an \emph{antipodal duality} between this form factor and certain six-particle amplitudes in the same theory~\cite{Dixon:2021tdw}. Explicitly, this duality relates the six-particle amplitude, evaluated on the parity-conserving kinematic surface (namely, in kinematics that are left invariant by the action of spacetime parity), to the \emph{polylogarithmic antipode} of the three-particle form factor, once an appropriate mapping between the kinematic variables that describe these processes has been identified. Notably, the antipode map reverses the order of letters in the symbol, namely sending
\begin{align}
    x_1 \otimes x_2 \otimes \cdots \otimes x_n \to (-1)^n x_n \otimes \cdots \otimes x_2 \otimes x_1\, .
\end{align}
You might wonder what the physical interpretation of the antipode map is (or even of the full antipodal duality)---but at the moment the answer is \emph{we do not know}. This duality remains an observation, although it has been checked to hold analytically through eight loops (an extremely nontrivial check, looking back at Table~\ref{tab:form-factor-symbol}). Among other things, this implies that the discontinuities of the amplitude are encoded in the derivatives of the form factor, and vice versa. This duality is conjectured to hold at all loop orders (and even has extensions to higher-particle form factors~\cite{Dixon:2022xqh}), but at the moment has no compelling physical interpretation. Thus, more bootstrap data is needed!

\begin{acknowledgements}
	-- The \href{https://indico.sns.it/event/91/}{{\it La Ricotta} Summer School  {\it The Disordered Universe 2025}} received financial support from the ERC grant NOTIMEFORCOSMO (agreement n${}^{\circ}$ 1126304)), by Scuola Normale and by the Instituto Galego de F{\'i}sica das Altas Enerx{\'i}as (IGFAE-USC/Xunta de Galicia) through the CIGUS network and EU cofinancing program.
\end{acknowledgements}

\newpage
\appendix
\section{Exercises}\label{app:exer}

\subsection{Tutorial 1}
\label{sec: tutorial 1}
\begin{exbox}
\begin{enumerate}
\item In this problem, we're going to study how singularities can ``move around'' in different integral representations. We'll use the example of the massless box Feynman integral:
\begin{center}
\resizebox{3.6cm}{!}{\begin{tikzpicture}[]
        \coordinate (a) at (-.8,.8);
        \coordinate (b) at (.8,.8);
        \coordinate (c) at (.8,-.8);
        \coordinate (d) at (-.8,-.8);
        \draw[] (a) to (b);
        \draw[] (b) to (c);
        \draw[] (c) to (d);
        \draw[] (d) to (a);
        \draw[] (a) --++ (135:1);
        \draw[] (b) --++ (45:1);
        \draw[] (c) --++ (-45:1);
        \draw[] (d) --++ (-135:1);
        
        \node[] at (-1.8,1.7) {$p_1$};
        \node[] at (-1.8,-1.7) {$p_2$};
        \node[] at (1.8,-1.7) {$p_3$};
        \node[] at (1.8,1.7) {$p_4$};

        \node at ($(a)!0.5!(b)$) [above]{$\alpha_4$};
        \node at ($(a)!0.5!(d)$) [left]{$\alpha_1$};
        \node at ($(b)!0.5!(c)$) [right]{$\alpha_3$};
        \node at ($(c)!0.5!(d)$) [below]{$\alpha_2$};
              \end{tikzpicture}}
   \end{center}

\begin{enumerate}
\item\label{ex:1-1a} Write down and then solve the Landau equations in momentum space, for the solution in which $\alpha_2 = \alpha_4 = 0$ but $\alpha_3 \neq 0 \neq \alpha_1$.
\item\label{ex:1-1b} What is the dimension of the solution you found in the space of Feynman parameters? Will this singularity appear on the physical Riemann sheet?
\item\label{ex:1-1c} The Lee-Pomeransky representation of the box integral is given by
\begin{equation*}
\sim \int_{\alpha_i \ge 0} \frac{ \dd \alpha_1 \dd \alpha_2 \dd \alpha_3 \dd \alpha_4}{\mathcal{G}^{D/2}}
\end{equation*} 
where $D$ is the spacetime dimension, $\mathcal{G} = \mathcal{F} + \mathcal{U}$, and
\begin{equation*}
\mathcal{F} = -s \alpha_2 \alpha_4 - t \alpha_1 \alpha_3 \, , \qquad \mathcal{U} = \alpha_1 + \alpha_2 + \alpha_3 + \alpha_4
\end{equation*}
in terms of the Mandelstam invariants $s = (p_1+p_2)^2$, $t= (p_2 + p_3)^2$. Derive the Landau equations for this representation of the box integral.
\item\label{ex:1-1d} Solve the Lee-Pomeransky version of the Landau equations for the situation in which all four $\alpha_i \neq 0$.
\item\label{ex:1-1e} Now solve the Lee-Pomeransky Landau equations when $\alpha_2 = \alpha_4 = 0$ but $\alpha_3 \neq 0 \neq \alpha_1$. What happened to the solution we found in momentum space?
\end{enumerate}
It turns out we need to carry out an algebraic blowup to recover the solution we found before in momentum space. In this case, we rescale our variables via
\begin{equation*}
\alpha_i \to \lambda \alpha_i
\end{equation*}
and approach a specific corner of the integration region by taking the $\lambda \to \infty$ limit. 
\begin{enumerate}
\item[(f)]\label{ex:1-1f} What does the $\mathcal{G}$ polynomial reduce to at leading order as $\lambda \to \infty$? If we solve the Landau equations just for this leading coefficient (again setting $\alpha_2 = \alpha_4 = 0$ and assuming $\alpha_3 \neq 0 \neq \alpha_1$), do we recover the solution we found in momentum space?
\end{enumerate}

\newpage
\item Now let's see how we can predict the \emph{nature} of the singularities that arise in an integral. Consider again the toy example:
\begin{equation*}
\int_{-1}^1 \frac{\dd x}{(x-z)(x+z)} 
\end{equation*}
The singularity at $z=0$ in this integral arises at an isolated point in $x$ space.
\begin{enumerate}
\item\label{ex:1-2a} We can approximate the behavior of this integral near $z=0$ by integrating over just the region within some distance $\delta \ll 1$ of where the singularity arises, and then (keeping $\delta$ fixed) taking the $z \to 0$ limit. What type of dependence on $z$ does this predict? (Note that $\delta$ should drop out.) Feel free to use the following result for the indefinite integral:
\begin{equation*}
\int \frac{1}{(x-z)(x+z)} \dd x = \frac{\log(1-x/z) - \log(1+x/z)}{2 z} \, .
\end{equation*}
\end{enumerate}
Now let's consider the massive bubble integral
\begin{center}
\begin{tikzpicture}[baseline]
    \draw (-3, 0) -- (-1, 0);
    \node at (-2, 0.6) {$p$};
    \draw[-{Latex[scale=1.5]}] (-2.5,0.3) -- (-1.5,0.3);
    \draw (1, 0) -- (3, 0);
    \draw (0,0) circle (1);
    \node at (0,1.25) {$m_1$};
    \node at (0, -1.25) {$m_2$};

\end{tikzpicture}
\end{center}
In exactly two spacetime dimensions, we saw that this integral can be written as a one-fold Feynman parameter integral
\begin{equation*}
\int_0^1 \frac{\dd \alpha}{s \alpha (1-\alpha) - m_1^2 \alpha - m_2^2 (1-\alpha)}
\end{equation*} 
If we solve for the roots of the denominator, this integral can be put in the form 
\begin{equation*}
\int_0^1 \frac{\dd \alpha}{(\alpha - A - \sqrt{B})(\alpha - A + \sqrt{B})}
\end{equation*} 
where $A$ and $B$ are rational functions of $s$, $m_1^2$, and $m_2^2$. In the threshold and pseudo-threshold limits, $B \to 0$ while $A$ remains finite.
\begin{enumerate}
\item[(b)]\label{ex:1-2b} Using the same method as before, predict the type of behavior your expect in the threshold and pseudo-threshold limits. (You can assume we've analytically continued to a region in which the pseudo-threshold singularity does indeed arise.)
\item[(c)]\label{ex:1-2c} How general is this approximation strategy? When do you expect it to break down?
\end{enumerate}
\end{enumerate}
\end{exbox}

\subsection{Tutorial 2}

\begin{exbox}
\begin{enumerate}
\item Imagine we've shown that the integral $I$ obeys the hierarchical constraint \label{ex:2-1}
\begin{equation*}
\text{Disc}_{t-4m^2} \cdots \text{Disc}_{s-t} \cdots (I) = 0\, .
\end{equation*}
Which of the following symbol terms (some of which have algebraic prefactors) are consistent with this constraint?
\begin{gather*}
\left( t-4m^2 \right) \otimes \left(s-t \right) \otimes \left ( s-4m^2 \right) \\[.2cm]
\left( \frac{\sqrt{t} - \sqrt{s}}{\sqrt{t}+ \sqrt{s}} \right) \otimes  \left( t-4m^2 \right)  \\[.2cm]
\left( \frac{m - \sqrt{s - t}}{m+ \sqrt{s - t}} \right) \otimes  \left( \frac{t-4m^2}{s} \right) \otimes \left( \frac{m - \sqrt{s - t}}{m+ \sqrt{s - t}} \right) \\[.2cm]
\left( t-4m^2 \right) \otimes \left( \frac{m - \sqrt{s - t}}{m+ \sqrt{s - t}} \right) \\[.2cm]
\frac{1}{\sqrt{s - t}}\left( \left( s-4m^2 \right) \otimes \left( t-4m^2 \right) \right) \\[.2cm]
\frac{1}{\sqrt{s - t}}\left( \left( t-4m^2 \right) \otimes \left( \frac{m - \sqrt{s - t}}{m+ \sqrt{s - t}} \right) \right)
\end{gather*}
Similarly, which of the following symbol terms are consistent with the Steinmann relations for the pair of thresholds $s=4 m^2$ and $t=4m^2$?
\begin{gather*}
\left( t-4m^2 \right) \otimes \left(t-4m^2 \right) \otimes \left ( s-4m^2 \right) \\[.2cm]
\left( t-4m^2 \right) \otimes \left(s-t \right) \otimes \left ( s-4m^2 \right) \\[.2cm]
\left( \frac{\sqrt{s-4m^2} - \sqrt{t-4m^2}}{\sqrt{s-4m^2} + \sqrt{t-4m^2}} \right) \otimes  \left(s-t \right) \otimes \left ( s-4m^2 \right) \\[.2cm]
\left( \frac{\sqrt{5} \, m - \sqrt{t-4m^2}}{\sqrt{5}\, m + \sqrt{t-4m^2}} \right) \otimes  \left(s-t \right) \otimes \left( \frac{\sqrt{5}\,m - \sqrt{s-4m^2}}{\sqrt{5}\,m + \sqrt{s-4m^2}} \right) \\[.2cm]
\left( \frac{\sqrt{5}\,m - \sqrt{s-4m^2}}{\sqrt{5}\,m + \sqrt{s-4m^2}} \right) \otimes \left ( s-4m^2 \right) \otimes \left( \frac{\sqrt{5}\,m - \sqrt{t-4m^2}}{\sqrt{5}\,m + \sqrt{t-4m^2}} \right) \otimes \left( \frac{\sqrt{5}\,m - \sqrt{t-4m^2}}{\sqrt{5}\,m + \sqrt{t-4m^2}} \right) \\[.2cm]
\frac{1}{\sqrt{s-4m^2}\sqrt{t-4m^2}} \left(\left( \frac{\sqrt{5}\,m - \sqrt{t-4m^2}}{\sqrt{5}\,m + \sqrt{t-4m^2}} \right) \otimes  \left(s-t \right) \otimes \left( \frac{\sqrt{5}\,m - \sqrt{s-4m^2}}{\sqrt{5}\,m + \sqrt{s-4m^2}} \right) \right) 
\end{gather*}

\newpage
\item\label{ex:2-2}
Just using logs and square roots, see if you can construct a function $F(y,z)$ that satisfies the constraint
\begin{equation*}
\text{Disc}_y (F(y,z)) = \text{Disc}_y ( \text{Disc}_z  (F(y,z))) \, .
\end{equation*} 
What about a function in which $y$ and $z$ only enter via the square roots $\sqrt{y}$ and $\sqrt{y-z}$ (possibly also as the arguments of logarithms)?
\vspace{1cm}

\item Consider the four-fold integral 
\begin{equation*}
\int \frac{\dd x_1 \dd x_2 \dd y_1 \dd y_2}{(x_1 - x_2 + y_1 + C)(-x_1 + x_2 + y_2 + C)(x_1 x_2 + y_1^2  y_2^2)} \, ,
\end{equation*}
where $C$ is a constant.

\begin{enumerate}
\item\label{ex:2-3a} What type of special functions will arise if you just evaluate the integrals over $x_1$ and $x_2$? \\[-.2cm]
\item\label{ex:2-3b} What type of special functions will arise if you just evaluate the integrals over $y_1$ and $y_2$?\\[-.2cm]
\item\label{ex:2-3c} What type of special functions will arise if you evaluate all four integrations? \\[-.2cm]
\item\label{ex:2-3d}  Finally, what type of special functions will arise if one first sets $C=0$ before evaluating all four integrations?
\end{enumerate}
\end{enumerate}
\end{exbox}

\newpage
\section{Solutions}\label{app:soln}

\subsection{Tutorial 1}

\begin{itemize}
    \item[(\ref{ex:1-1a})] Using $\alpha_2=\alpha_4=0$, the Landau equations \eqref{eq:landua1} and \eqref{eq:landua2} imply that \begin{gather}
        q_1^2=q_3^2=0 \, ,\\ 
        \alpha_1q_1^\mu+\alpha_3q_3^\mu=0 \, .
    \end{gather}
    Parameterizing our loop momentum such that $q_1^\mu=\ell^\mu$, these equations become
    \begin{gather}
        \ell^2=(\ell+p_2+p_3)^2=0  \\ 
        \alpha_1\ell_1^\mu+\alpha_3 \, ,(\ell^\mu+p_2^\mu+p_3^\mu)=0\, .
    \end{gather}
    Solving the second equation for the loop momentum, we find that $\ell^\mu = -\frac{\alpha_3}{\alpha_1+\alpha_3} \left( p_2^\mu + p_3^\mu \right) $.
    We thus have that $t = (p_2+p_3)^2 \propto \ell^2 = 0$. The full solution to the Landau equations thus reads 
    \begin{gather}
        \alpha_2=\alpha_4=0 \, ,\\
        \alpha_1+\alpha_3=1 \, ,\\
        \ell^2=\ell\cdot(p_2+p_3)=t=0\, .
    \end{gather}
    This is a threshold singularity.
    \item[(\ref{ex:1-1b})] This solution is one-dimensional in the space of Feynman parameters. Moreover, it intersects the regions where all $\alpha_i \ge 0$, so it is $\alpha$-positive and will appear on the physical Riemann sheet.
    \item[(\ref{ex:1-1c})] In the Lee-Pomeransky representation, the Landau equations are given by $\mathcal{G} = 0$ and $\alpha_i \left(\frac{\partial\mathcal{G}}{\partial\alpha_i} \right)  = 0$ for $i \in \{1,2,3,4 \}$. In this example, these equations become
\begin{gather}    
        \mathcal{G} = -s \alpha_2 \alpha_4 - t   \alpha_1 \alpha_3  + \alpha_1 + \alpha_2 + \alpha_3 + \alpha_4 =0   \, ,   \\
     \alpha_1 \left(\frac{\partial\mathcal{G}}{\partial\alpha_1}\right) = \alpha_1 (1-t \alpha_3) = 0\, , \\
      \alpha_2\left(\frac{\partial\mathcal{G}}{\partial\alpha_2}\right) = \alpha_2(1-s \alpha_4) = 0\, , \\
      \alpha_3 \left(\frac{\partial\mathcal{G}}{\partial\alpha_3}\right) = \alpha_3 (1-t \alpha_1) = 0\, , \\
      \alpha_4 \left(\frac{\partial\mathcal{G}}{\partial\alpha_4}\right) = \alpha_4 (1-s \alpha_2) = 0 \, .
\end{gather}
    \item[(\ref{ex:1-1d})] When all of the $\alpha_i$ parameters are nonzero, the Landau equations are solved by $\alpha_1=\alpha_3=1/t$, $\alpha_2=\alpha_4=1/s$. Substituting this into $\mathcal{G}=0$, we find $s+t=0$.
    \item[(\ref{ex:1-1e})] We now have the reduced set of equations 
\begin{gather}   
    - t   \alpha_1 \alpha_3 + \alpha_1 + \alpha_3 =0 \, , \\  
    1-t \alpha_3 = 0\, , \\
    1-t \alpha_1 = 0\, .
\end{gather}    
From the last two equations, we see $\alpha_1=\alpha_3=1/t$, which---when substituted into the first equation---yields $1/t=0$. This is no longer the same solution we found in part (a)!
    \item[({\protect\hyperref[ex:1-1f]{1f}})] After rescaling $\alpha_i\to\lambda\alpha_i$, we have $\mathcal{G}=\lambda^2 \mathcal{F}+\lambda\mathcal{U}$. To leading order around $\lambda\to\infty$, we see that the $\mathcal{F}$ polynomial dominates, namely $\mathcal{G}\to\mathcal{F}=-s\alpha_2\alpha_4-t\alpha_1\alpha_3$. Thus, solving the Landau equations at leading order in this limit, and using the assumptions that $\alpha_2=\alpha_4=0$ while $\alpha_1\neq0\neq\alpha_3$, we recover the solution we found in part (a).
    \item[(\ref{ex:1-2a})] Using the given indefinite integral, we find
    \begin{align*}
        \int_{-\delta}^\delta\frac{\dd x}{(x-z)(x+z)} &= \frac{\log(1-\delta/z)-\log(1+\delta/z)-\log(1+\delta/z)+\log(1-\delta/z)}{2z} \\ 
        &= \frac{1}{z}\log\frac{z-\delta}{z+\delta}.
    \end{align*}
    As $z\to 0$, this integral thus behaves as $i \pi/z$. 
    \item[({\protect\hyperref[ex:1-2b]{2b}})] We make the substitution $\beta=\alpha-A$, such that the integral becomes
    \begin{align*}
        \int_{-A}^{1-A} \frac{\dd \beta}{(\beta-\sqrt{B})(\beta+\sqrt{B})}.
    \end{align*}
    This is functionally equivalent to the previous exercise, and hence as $B\to 0$ this integral behaves as $2\pi i/\sqrt{B}$.
    \item[({\protect\hyperref[ex:1-2c]{2c}})] It could be the case that multiple singularities develop along the integration contour when we approach a given point in the space of external kinematics. To get a reliable prediction, we have to make sure we are taking all singularities into account. Also, we need to be able to evaluate the integral near each of these singularities.
\end{itemize}

\subsection{Tutorial 2}

\begin{itemize} 
    \item[(\ref{ex:2-1})] The first and last symbols are consistent with the hierarchical prediction, but the others are not. In the first symbol, the letter $t-4m^2$ only appears before the letter $s-t$, so there is no chance to compute a discontinuity with respect to the former after computing a discontinuity with respect to the latter. In the second symbol, the first letter encodes a logarithmic discontinuity at $s=t$, after which we still have the disallowed letter $t-4m^2$. The third symbol is a trickier one; we can first compute a logarithmic discontinuity with respect to $s - t - m^2$, after which the remaining symbol is odd under the sign flip $\sqrt{s-t} \to -\sqrt{s-t}$, indicating we can compute an algebraic discontinuity with respect to $s-t$; after this, we still have a logarithmic discontinuity that we can access at $t-4m^2$. The fourth symbol violates the hierarchical prediction for the same reason as the third. In the fifth symbol, we can compute an algebraic discontinuity with respect to $s-t$, and then a logarithmic discontinuity with respect to $t-4m^2$. However, in the sixth symbol, we cannot do this because the overall symbol is even under the sign flip  $\sqrt{s-t} \to -\sqrt{s-t}$. The only way to uncover a discontinuity with respect to $s-t$ is to get rid of the second letter (by computing two logarithmic discontinuities) in order to make the symbol odd under $\sqrt{s-t} \to -\sqrt{s-t}$; however, by this point there is no longer any branch point at $t=4m^2$.

The second and fifth symbols satisfy the Steinmann relations, but the others do not. The first symbol has the disallowed pair of letters next to each other in the second and third entry. In the second symbol, however, the restricted letters are separated by the letter $(s-t)$, so the symbol term is allowed. In the third symbol, the first letter encodes nonzero discontinuities with respect to both $s-4m^2$ and $t-4m^2$, neither of which is logarithmic, so Steinmann is violated. The fourth symbol also has a pair of algebraic discontinuities with respect to these thresholds (now in the first and third letters, but all that matters is whether the symbol term is even or odd with respect to flipping the sign in front of these square roots), leading again to Steinmann violation. Conversely, in the fifth symbol the first two letters encode discontinuities with respect to $s-4m^2$, but since the symbol is even with respect to $\sqrt{t-4m^2} \to -\sqrt{t-4m^2}$ we cannot compute a nonzero discontinuity with respect to $t-4m^2$ without first computing at least three logarithmic discontinuities (in order to make the symbol odd under $\sqrt{t-4m^2} \to -\sqrt{t-4m^2}$); however, getting rid of the third letter requires computing a discontinuity with respect to $t-9 m^2$, and after computing this discontinuity there are no $s-4m^2$ discontinuities left. Finally, the last symbol does not satisfy Steinmann because we can always compute a maximal number of logarithmic discontinuities, leaving just the algebraic prefactor that has algebraic discontinuities at both $s=4m^2$ and $t=4m^2$.

\item[(\ref{ex:2-2})] The massive bubble integral in $D=2$ has this property, when $z = m_1^2$ and $y = p^2-(m_1+m_2)^2$.
\item[(\ref{ex:2-3a})] We start with an integrand that has three denominator factors, each of which is (separately) linear in $x_1$ and $x_2$. Computing any of the possible residues with respect to either $x_1$ or $x_2$, we find that the resulting denominator factors are still all polynomials of the remaining integration variables. This means we can always compute a further residue with respect to the remaining integration variable, so the integral will evaluate to a multiple polylogarithm.
\item[(\ref{ex:2-3b})] Similar to part (3a), after computing any of the possible residues with respect to either $y_1$ or $y_2$, we are left with denominator factors that are polynomials of the remaining variables, so the integral will evaluate to a multiple polylogarithm.
\item[(\ref{ex:2-3c})] We need to check all possible ways of computing a maximal sequence of residues, and analyze any obstructions that stop us from computing residues with respect to all four integration variables. For instance, if we compute a residue of $x_1$ at $x_2-y_1-C$, we get
\begin{equation}
        \frac{1}{(y_1+y_2+2C)(x_2^2-x_2y_1-Cx_2+y_1^2y_2^2)}\,.
\end{equation}
Computing a second residue of $y_2$ at $-y_1-2C$, we find 
\begin{equation}
        \frac{1}{-C x_2 + x_2^2 - x_2 y_1 + 4 C^2 y_1^2 + 4 C y_1^3 + y_1^4}\,.
\end{equation}
Since this denominator factor is still just quadratic in $x_2$, we can easily compute another residue with respect to one of its two $x_2$ roots. The result will be
\begin{equation}
         \frac{\pm1}{\sqrt{C^2 + 2 C y_1 + y_1^2 - 16 C^2 y_1^2 - 16 C y_1^3 - 4 y_1^4}}\,, \label{eq:elliptic_obstruction}
\end{equation}
where the overall sign depends on which root we have computed a residue with respect to. By computing the $j$-invariant of the quartic polynomial of $y_1$ that appears in this square root, we find that the integrand that remains involves a non-factorizable elliptic curve. Thus, we conclude that this four-fold integral evaluates to an iterated integral that depends on this elliptic curve. Do we expect any other irreducible curves to appear in the answer? 
\item[(\ref{ex:2-3d})] After setting $C=0$, we can carry out the same type of analysis as we did in the previous parts of this problem. Alternately, we can check that in the $C\to0$ limit, the curves that we identified in our previous analysis degenerate. In particular, we can check that the elliptic integrand from~\eqref{eq:elliptic_obstruction} simplifies, becoming
\begin{equation}
         \frac{\pm1}{\sqrt{y_1^2 (1 - 2 y_1) (1 + 2 y_1)}} = \frac{\pm1}{y_1 \sqrt{(1 - 2 y_1) (1 + 2 y_1)}}\,.
\end{equation}
Thus, we can compute one further residue than we could before, so this elliptic obstruction evaporates.
\end{itemize}

\bibliographystyle{utphys}
\bibliography{2025-DU-AnStr-biblio}

\end{document}